\documentstyle[12pt]{article}
\catcode`@=11
\newcount\@tempcntc
\def\@citex[#1]#2{\if@filesw\immediate\write\@auxout{\string\citation{#2}}\fi
  \@tempcnta\z@\@tempcntb\m@ne\def\@citea{}\@cite{\@for\@citeb:=#2\do
    {\@ifundefined
       {b@\@citeb}{\@citeo\@tempcntb\m@ne\@citea\def\@citea{,}{\bf ?}\@warning
       {Citation `\@citeb' on page \thepage \space undefined}}%
    {\setbox\z@\hbox{\global\@tempcntc0\csname b@\@citeb\endcsname\relax}%
     \ifnum\@tempcntc=\z@ \@citeo\@tempcntb\m@ne
       \@citea\def\@citea{,}\hbox{\csname b@\@citeb\endcsname}%
     \else
      \advance\@tempcntb\@ne
      \ifnum\@tempcntb=\@tempcntc
      \else\advance\@tempcntb\m@ne\@citeo
      \@tempcnta\@tempcntc\@tempcntb\@tempcntc\fi\fi}}\@citeo}{#1}}
\def\@citeo{\ifnum\@tempcnta>\@tempcntb\else\@citea\def\@citea{,}%
  \ifnum\@tempcnta=\@tempcntb\the\@tempcnta\else
   {\advance\@tempcnta\@ne\ifnum\@tempcnta=\@tempcntb \else \def\@citea{--}\fi
    \advance\@tempcnta\m@ne\the\@tempcnta\@citea\the\@tempcntb}\fi\fi}
\catcode`@=12

\def\theequation{\arabic{section}.\arabic{equation}}
\begin{document}

\begin{flushright}
NYU-TH/95-03\\[-0.2cm]
RAL-TR/95-022\\[-0.2cm]
July 1995
\end{flushright}

\begin{center}
{\bf{\LARGE A Gauge-Independent Approach to}}\\[0.4cm]
{\bf{\LARGE Resonant Transition Amplitudes}}\\[2.4cm]
{\large Joannis Papavassiliou}$^a$
{\large and Apostolos Pilaftsis}$^b$\footnote[1]{E-mail address:
pilaftsis@v2.rl.ac.uk}\\[0.4cm]
$^a${\em Department of Physics, New York University, 4 Washington Place,
New York,\\
NY 10003, USA}\\[0.3cm]
$^b${\em Rutherford Appleton Laboratory, Chilton, Didcot, Oxon, OX11 0QX, UK}
\end{center}
\vskip1.5cm
\centerline {\bf ABSTRACT}

\noindent
We present a new gauge-independent approach to resonant transition
amplitudes with nonconserved external currents, based
on the pinch technique method. In the context of $2\to 2$ and
$2\to 3$ scattering processes, we show explicitly that the analytic results
derived respect $U(1)_{em}$ gauge symmetry and do not depend on the choice
of the $SU(2)_L$ gauge fixing. Our analytic
approach treats, on equal footing, fermionic as well as bosonic
contributions to the resummed gauge boson propagators, does not contain
any residual space-like threshold terms, shows the correct high-energy
unitarity behaviour, admits renormalization, and satisfies a number of
other required properties, including the optical theorem. Even though
our analysis has mainly
focused on the Standard Model gauge bosons,
our method can easily be extended to the top quark,
and be directly applied to the study of
unstable particles present in renormalizable
models of new physics.\\[0.3cm]

\newpage

\section{Introduction}
\indent

Several years after the first experimental observations of decaying
quantum mechanical systems~\cite{RS}, Weisskopf and
Wigner~\cite{WWW} formulated a theory for the time evolution
of decaying states, which has been used with great success for
the description of CP violation in the $K_0-\bar{K}_0$ and
other systems.
This theory is however approximate, and deviations
from its predictions are expected, when observations
take place at very short or very long times as compared to the
lifetime of the unstable particle~\cite{JSLK}. Subsequently,
Veltman~\cite{Velt} showed that an $S$-matrix theory,
where the dynamics of unstable particles is described
in terms of initial and final asymptotic states,
is unitary and causal,
despite the presence of on-shell particle configurations.

The correct treatment of unstable particles
has received a renewed attention within the framework of the $S$-matrix
perturbation theory, mainly because the straightforward generalization
of the Breit-Wigner (BW) propagator derived
from naive scalar field theories~\cite{Velt}
to gauge field theories, violates the gauge
symmetry~\cite{AP,WV,RGS,AS,NP,CLP,HV,AOW,BZ}.
This fact is perhaps not so surprising, since the
naive resummation of
the self-energy graphs takes into account higher order corrections, for
{\em only} certain parts of the tree-level amplitude.
Even though, as we will show, the amplitude possesses all the
desired properties,
this unequal treatment of its parts
distorts subtle cancellations,
resulting in numerous pathologies, which are artifacts of the
method used. Evidently,
a self-consistent calculational scheme needs be devised,
which will exploit
all the healthy field theoretical properties intrinsic in
every $S$-matrix element.

An early attempt in this direction has been based on the observation
that the position of the complex pole
is a gauge independent (g.i.)
quantity~\cite{WV,RGS,AS}. Exploiting this
fundamental property of the $S$-matrix,
Stuart~\cite{RGS} has developed a perturbative
approach in terms of three gauge invariant quantities:
the constant complex pole position of the resonant amplitude, the residue
of the pole, and a $q^2$-dependent non-resonant background term.
Even though this approach,
which is based on a Laurent series expansion of the
resonant transition element~\cite{RGS}, may
eventually furnish a gauge invariant result,
the perturbative treatment of these three g.i.\
quantities~\cite{HV} introduces unavoidably residual space-like
threshold terms, which become more apparent in CP-violating
scenarios of new-physics. In fact, the precise $q^2$-dependent shape
of a resonance~\cite{AS} is reproduced, to a given loop order, by considering
quantum corrections to the three g.i.\ quantities mentioned
above~\cite{RGS,HV}, while the space-like threshold contributions,
even though
are shifted to higher orders, do not disappear completely.

Within the framework of the $S$-matrix perturbation theory,
it was suggested~\cite{AP} that finite width effects can induce sizeable
CP violation and {\em resonantly} enhance CP-violating observables~\cite{PN}
in supersymmetric theories, and other extensions of the minimal
Standard Model (SM)~\cite{CPviol}. The quest of the proper BW
form for a resonant $W$ and $t$ propagator~\cite{NP,CLP,AESMM}
is equally important for processes, such as $e^+e^-\to W^+W^-$~\cite{AOW},
$e^-\gamma\to \mu^-\bar{\nu}_\mu\nu_e$~\cite{BZ,BVZ}, {\em etc.}

In this paper, we present a new g.i.\ approach
to resonant transition amplitudes implemented
by the pinch technique (PT)~\cite{JMC,JPself,JPgeneral,JPKP}.
The PT is an algorithm
that systematically exploits the known field theoretical
properties of the $S$-matrix,
which is the fundamental physical quantity of interest.
Operationally, the PT
leads to a
rearrangement of the Feynman graphs contributing to
a gauge-invariant amplitude,
in such a way
as to define
individually g.i.\ propagator,vertex, and
box-like structures.
For example, the PT arranges the $S$-matrix
element $T$ for the process
$q_{1}{\bar{q}}_{2}\rightarrow q_{1}{\bar{q}}_{2}$,
where $q_{1}$, $q_{2}$ are two
on-shell test quarks with masses
$m_{1}$ and $m_{2}$, in the form
\begin{equation}
T(s,t,m_{1},m_{2})=
{\widehat{T}}_{1}(t) + {\widehat{T}}_{2}(t,m_{1})+
{\widehat{T}}_{2}(t,m_{2})+
{\widehat{T}}_{3}(s,t,m_{1},m_{2})~,
\label{S2-matrix}
\end{equation}
where the ${\widehat{T}}_{i}$ ($i=1,2,3$) are {\em individually}
$\xi$ independent.
  The parts of vertex and box graphs which
are kinematically akin to propagators and
enforce the gauge independence of ${\widehat{T}}_{1}(t)$,
are called propagator-like pinch parts. Similarly,
vertex-like pinch parts of boxes enforce the
gauge independence of $T_{2}(t)$.

The crucial novel ingredient we introduce
in the context of resonant transition amplitudes
is the proposition that
the resummation of graphs must take place only
{\em after} the amplitude of interest has been cast
via the PT algorithm into
manifestly g.i.\ sub-amplitudes, with distinct
kinematic properties, order by order in perturbation theory.
For example, it is the resummations of
the ${\widehat{T}}_{1}$ which will provide the effective, manifestly
g.i., resummed propagators.

The main points of our approach have already presented
in a brief communication \cite{JP&AP}; in this paper
we mainly focus on the detailed treatment of several technical issues.
The outline of the present work is as follows.
In Section 2, we define the framework of our perturbative
g.i.\ $S$-matrix approach by considering the resonant
reaction $e^-\bar{\nu}_e\to \mu^-\bar{\nu}_\mu$. Issues of resummation
and the resummation procedure within the PT will be discussed in Section 3
and 4, respectively. In Section~5, we show that the position of the pole
does not get shifted when using the PT resummation algorithm in the stable
particle theory ---a heuristic proof is given in Appendix A. In Section~6,
we further show that this is still true for the case of unstable particles.
Section~7 deals with issues related to unitarity of resonant processes.
In Section 8, we give an application of our approach to the resonant
processes $\gamma e^-\to \mu^-\bar{\nu}_\mu\nu_e$ and $QQ'\to e^-\bar{\nu}_e
\mu^-\mu^+$, which involve the $\gamma WW$ and $ZWW$ vertices, respectively.
Further technical details of such reactions are relegated in Appendices B
and C. Section 9 contains our conclusions.

\setcounter{equation}{0}
\section{The process $e^-\bar{\nu}_e\to \mu^-\bar{\nu}_\mu$}
\indent

Despite the fact that the $S$ matrix is well defined, the
evaluation of physical processes has to rely on its perturbative
expansion in the coupling constants of the theory, as there
is not yet an analytic method to calculate the complete $S$-matrix
amplitude. On the other hand, this perturbative
approximation of $S$ is not unique, and depends on the form
of the expansion adopted, and, to some extend, on the renormalization
prescription used to remove the ultra-violet (UV) divergences. However,
the summation of all infinite perturbative contributions should formally
reproduce the unique expression of the $S$-matrix element of the process
under consideration. Although the perturbative expansion itself
may contain such difficulties, there are some well-defined features
that characterize a consistent perturbative expansion of $S$ matrix within
gauge field theories:
\begin{itemize}
\item[(i)] The expansion should obey a number of required properties, including
unitarity [or equivalently the optical theorem]~\cite{Velt},
causality~\cite{LSZBS}, analyticity
{\em etc.}~\cite{ELOP}

\item[(ii)] Since we are interested in renormalizable field theories
based on Lagrangians which contain operators of dimension no higher
than four and so have an inherent predictive power,
the expansion under consideration should consistently admit renormalization.

\item[(iii)]
The perturbative $S$-matrix element should respect the fundamental gauge
symmetries. In particular, since it represents a physical quantity,
it should be independent on the choice of gauge used, which can only be
shown to be the case with the help of Becchi--Rouet--Stora (BRS)
transformations~\cite{BRS}.
\end{itemize}

Conditions (i) and (iii) are the main source of problems,
when considering resonant $S$-matrix transition amplitudes.
In what follows, we will discuss some of the crucial differences
between our approach and the conventional $S$-matrix perturbation theory.
In the context of the latter,
the one-loop $W$-boson self-energy has the general form
\begin{equation}
\Pi_{\mu\nu}^{(\xi )}(q)\ =\ t_{\mu\nu}(q)
\Pi_T^{(\xi )}(q^2)\,
+\, \ell_{\mu\nu}(q) \Pi_L^{(\xi )}(q^2), \label{Pixi}
\label{TL}
\end{equation}
where
\begin{eqnarray}
t_{\mu\nu}(q) &=& -\, g_{\mu\nu}\, +\, \frac{q_\mu q_\nu}{q^2},
\nonumber\\
\ell_{\mu\nu}(q) &=& \frac{q_\mu q_\nu}{q^2}.
\end{eqnarray}
The self-energy of
Eq.~(\ref{Pixi}) is a gauge-dependent quantity;
in the conventional $S$-matrix approach it depends
explicitly on the
gauge parameter $\xi$.
The two-point function for the mixing $W^-G^-$, $\Theta_\mu$,
and $G^-G^-$ self-energy, $\Omega$, are also $\xi$-dependent.
Using the general form of Eq.~(\ref{Pixi}) for the self-energy,
the one-loop resummed $W$ propagator is given by
\begin{eqnarray}
\Delta^{(\xi)}_{\mu\nu} (q) &=& \left( \Delta_{0\mu\nu}^{-1(\xi)}
(q)\, -\, \Pi_{\mu\nu}^{(\xi )}(q) \right)^{-1}\nonumber\\
 &=& t_{\mu\nu}(q)\, \frac{1}{q^2\, -\, M^2\, -\,
\Pi_T^{(\xi )}(q^2)}\ -\ \ell_{\mu\nu}(q)\, \frac{\xi}{q^2\ -\ \xi (
M^2- \Pi_L^{( \xi )}(q^2))}, \label{Deltaxi}
\end{eqnarray}
where
\begin{eqnarray}
\Delta_{0\mu\nu}^{(\xi)}(q) &=& t_{\mu\nu}(q)\, \frac{1}{q^2 -
M^2}\ -\ \ell_{\mu\nu}(q)\, \frac{\xi}{q^2-\xi M^2}\nonumber\\
&=& U_{\mu\nu}(q)\ -\ \frac{q_\mu q_\nu}{M^2} D_0^{(\xi)}(q^2)\, .
\label{D0xi}
\end{eqnarray}
In Eq.\ (\ref{D0xi}), $U_{\mu\nu}$ stands for the free $W$ propagator
in the unitary gauge, which has the form
\begin{eqnarray}
U_{\mu\nu}(q) &=& [-g_{\mu\nu}+\frac{q_\mu q_\nu}{M^2}]
\frac{1}{q^2 -M^2}\nonumber\\
&=& t_{\mu\nu}(q)\, \frac{1}{q^2 -
M^2}\ +\ \ell_{\mu\nu}(q)\, \frac{1}{M^2}\, ,
\label{U1}
\end{eqnarray}
and
\begin{equation}
D_0^{(\xi)}(q^2) = \frac{1}{q^2-\xi M^2}\, ,
\label{D0}
\end{equation}
is the tree-level propagator of the associated Goldstone boson
$G^{+}$ in a general $\xi$ gauge.
Its resummed propagator reads
\begin{equation}
D^{(\xi)}(q^2)\ =\ \frac{1}{q^2\, -\, \xi M^2\, -\,
\Omega^{(\xi )}(q^2)}\, .\label{Dxi}
\end{equation}
For purposes of illustration,
we have only considered the lowest order of
resummation, where higher order $W^-G^-$ mixing effects
have not been taken into account. However, our conclusions
will still be valid for the general case.
Using the resummed
$\xi$-dependent propagators given in Eqs.~(\ref{Deltaxi}) and (\ref{Dxi})
for the calculation of a resonant process, such as
$e^-\bar{\nu}_e\to \mu^-\bar{\nu}_\mu$, to a given order of
perturbation theory, one can then verify easily that
the $\xi$ dependence does not disappear. The reason is that
$\Pi_{\mu\nu}^{(\xi )}(q^2)$ is a $\xi$ dependent quantity
in a region not far away from the resonant point $q^2=M^2$
[only at this point the self-energy is g.i.] and the propagators
(\ref{Deltaxi}) and (\ref{Dxi}) induce $\xi$ dependence to
all orders, while $\xi$-dependent terms coming from vertices
and box graphs can remove this gauge dependence only to
a given order of the conventional perturbation theory.
Instead, within our framework, the above problems associated with the
resummed self-energies are absent, because the entire $\xi$ dependence
has been eliminated via the PT order by order in perturbation theory,
{\em before} resummation takes place.

We will now consider an approach implemented by the
PT. Within the PT framework, the transition
amplitude $T(s,t,m_i)$ of a $2\to 2$ process,
such as $e^-\bar{\nu}_e\to \mu^- \bar{\nu}_\mu$ shown in Fig.~1,
can be decomposed as
\begin{equation}
\label{TPT}
T(s,t,m_i)\ =\ \widehat{T}_1(s)\ +\ \widehat{T}_2(s,m_i)\ +\
\widehat{T}_3(s,t,m_i),
\end{equation}
in terms of three individually g.i.\ quantities:
a propagator-like part ($\widehat{T}_1$), a vertex-like piece
($\widehat{T}_2$),
and a part containing box graphs ($\widehat{T}_3$). The important observation
is that vertex and box graphs contain in general
pieces, which are kinematically akin to self-energy graphs
of the transition amplitude.
The PT is a systematic way of extracting such pieces and
appending them to the conventional self-energy graphs.
In the same way, effective gauge invariant
vertices may be constructed, if
after subtracting from the conventional vertices the
propagator-like pinch parts we add the vertex-like pieces coming from
boxes. The remaining purely box-like contributions are then
also g.i.
Finally, the entire $S$-matrix
can be rearranged in the form of Eq.~(\ref{TPT}).
In the specific example $e^-\bar{\nu}_e\to \mu^-\bar{\nu}_\mu$,
the piece $\widehat{T}_1$ consists of three
{\em individually} g.i.\ quantities:
The $WW$ self-energy $\widehat{\Pi}_{\mu\nu}$ (Fig.\ 1(a)), the $W^-G^-$ mixing
term $\widehat{\Theta}_{\mu}$ (Figs.\ 1(b) and 1(c)),\footnote[1]{
In fact, we define $\widehat{\Theta}_\mu(q)=\widehat{\Pi}^{W^- G^-}_\mu (q) =
\widehat{\Pi}^{G^- W^-}_\mu (q) = -\widehat{\Pi}^{W^+G^+}_\mu (q) =
-\widehat{\Pi}^{G^+W^+}_\mu (q)$,
where the momentum always flows from the left to the right in the language
of Feynman diagrams.}
and the $GG$ self-energy $\widehat{\Omega}$ (Fig.~1(d)).
Similarly, $\widehat{T}_2(s,m_i)$ consists of two pairs of
g.i.\ vertices $W e^-\bar{\nu}_e$, $G e^-\bar{\nu}_e$
(${\widehat{\Gamma}}_{\mu}^{(1)}$ and ${\hat{\Lambda}}^{(1)}$, given in
Figs.~1(e) and 1(f), respectively)
and $W\mu^-\bar{\nu}_\mu$ and $G\mu^-\bar{\nu}_\mu$
(${\widehat{\Gamma}}_{\mu}^{(2)}$ and
${\hat{\Lambda}}^{(2)}$, in Figs.~1(g)
and 1(h)).
In addition to being g.i.,
the PT self-energies and vertices possess a very crucial
property, {\em e.g.}\ they satisfy {\em tree-level} Ward identities,
summarized as follows:
\begin{eqnarray}
\label{PT1}
q^\mu q^\nu \widehat{\Pi}_{\mu\nu}\, -\, 2 Mq^\mu \widehat{\Theta}_\mu\, +\,
M^2
\widehat{\Omega} &=& 0\, ,\\
\label{PT2}
q^\mu\widehat{\Pi}_{\mu\nu}\, -\, M\widehat{\Theta}_\nu &=& 0\, , \\
\label{PT3}
q^\mu \widehat{\Theta}_\mu\, -\, M\widehat{\Omega} &=& 0\, ,\\
\label{PT4}
q^\mu \widehat{\Gamma}_{\mu}^{i}\, -\, M\hat{\Lambda}^{i} &=& 0\, , (i=1,2).
\end{eqnarray}
These Ward identities are a direct consequence of the requirement
that $\widehat{T}_{1}$ and $\widehat{T}_{2}$ are fully $\xi$ independent.
As explained in detail in \cite{JPself} and \cite{JPKP2}, after having
cancelled via the PT all $\xi$ dependences
inside loops, these Ward identities
enforce the final cancellations of the $\xi$ dependences stemming
from the tree-level propagators. In fact, the derivation
of the Ward identities does not
require knowledge of the closed expressions of the quantities involved.
To see how the final $\xi$
dependences cancel by virtue of the aforementioned
Ward identities we turn to $\widehat{T}_1$. After the PT process
has been completed, $\widehat{T}_1$ reads:
\begin{eqnarray}
\widehat{T}_1 &=& \Gamma_0^\sigma \Delta^{(\xi)}_{0\sigma\rho}\Gamma_0^\rho +
\Gamma_0^{\sigma}\Delta^{(\xi)}_{0\sigma\mu} \widehat{\Pi}^{\mu\nu}
\Delta^{(\xi)}_{0\nu\rho}\Gamma_0^\rho + \Lambda_0 D_0^{(\xi)} \Lambda_0
+ \Lambda_0 D_0^{(\xi)} \widehat{\Omega} D_0^{(\xi)}\Lambda_0\nonumber\\
&& +\Gamma_0^\sigma \Delta^{(\xi)}_{0\sigma\mu}\widehat{\Theta}^\mu
D_0^{(\xi)} \Lambda_0 + \Lambda_0 D_0^{(\xi)}\widehat{\Theta}^\nu
\Delta^{(\xi)}_{\nu\rho} \Gamma_0^\rho \nonumber\\
&=& \Gamma_0^\sigma U_{\sigma\rho} \Gamma_0^\rho
+ \Gamma_0^\sigma U_{\sigma\mu} \widehat{\Pi}^{\mu\nu}
U_{\nu\rho} \Gamma_0^\rho,
\label{qwer}
\end{eqnarray}
where in the second step the Ward identities
of Eqs.\ (\ref{PT2}) and (\ref{PT3}) were used.
Clearly, all $\xi$ dependence has disappeared.
We can actually go one step further and
rewrite this last $\xi$ independent expression as a sum of two
pieces, one transverse and one longitudinal,
by employing Eq.\ (\ref{U1}) and the Ward identities of
Eqs.\ (\ref{PT2}) and (\ref{PT3}).
Indeed, if we write ${\widehat{\Pi}}_{\mu\nu}$
in the form of Eq.\ (\ref{TL}), {\em i.e.}\
${\widehat{\Pi}}_{\mu\nu}= t_{\mu\nu}{\widehat{\Pi}}_{T}+
\ell_{\mu\nu} {\widehat{\Pi}}_{L}$
we have
\begin{eqnarray}
\widehat{\Pi}_T  &=& -\frac{1}{3}\, \left( \widehat{\Pi}^\sigma_\sigma -\
\frac{M^2}{q^2}\widehat{\Omega}\right)\, ,\\
\widehat{\Pi}_L &=& \frac{M^2}{q^2}\widehat{\Omega}\, ,
\end{eqnarray}
and so $\widehat{T}_1$ may be written as
\begin{equation}
\widehat{T}_1= \Gamma_0^\mu \left[\frac{t_{\mu\nu}}{q^{2}-M^{2}}
\left(\, 1+ \frac{{\widehat{\Pi}}_{T}}{q^{2}-M^{2}}\right)\, +\,
\frac{\ell_{\mu\nu}}{M^2}\left(\, 1+\frac{{\widehat{\Pi}}_{L}}{M^{2}}\right)
\right]\Gamma_0^\nu\, .
\label{T1L1}
\end{equation}

Let us now assume for a moment that the
 PT decomposition holds to any order in perturbation theory
(we will extensively discuss the validity of this assumption in
the next sections).
In such a case,
summing up contributions from all orders in perturbation
theory we obtain for $\widehat{T}_1$ (suppressing contraction
of Lorentz indices)
\begin{eqnarray}
\widehat{T}_1 &=& \Gamma_0 U \Gamma_0
+ \Gamma_0 U \widehat{\Pi}U \Gamma_0
+\Gamma_0 U \widehat{\Pi}U
\widehat{\Pi} U \Gamma_0 +\ \cdots\nonumber\\
&=& \Gamma_0 \hat{\Delta} \Gamma_0, \label{T1res}
\end{eqnarray}
with
\begin{equation}
\hat{\Delta}_{\mu\nu}(q)\ =\ t_{\mu\nu}(q)\, \frac{1}{q^2 -
M^2 - \widehat{\Pi}_T (q^2)}\ +\ \ell_{\mu\nu}(q)\,
\frac{1}{M^2-\widehat{\Pi}_L(q^2)}.
\label{Dres}
\end{equation}
It is important to emphasize that the propagator of
Eq.\ (\ref{T1res}) is {\it process-independent}; one arrives
at exactly the same expression for $\hat{\Delta}_{\mu\nu}$,
$\widehat{\Pi}_T $, and $\widehat{\Pi}_L $, regardless
of the quantum numbers of the external particles \cite{NJW}.
In the last step of Eq.\ (\ref{T1res}), we have assumed that
the analytic
continuation of the result to the resonant point $q^2=M^2$ will
not cause any theoretical difficulty.
In the case of the conventional propagator such an assumption is
justified, since
the resonant propagator can be directly derived as a solution of the
corresponding Dyson--Schwinger (DS) integral equation,
which is well defined,
even at the singular point $q^2=M^2$.
The reason is that the DS
integral equations can be deduced directly from the
action of the theory, through a variational principle \cite{CJT}.
Even though the corresponding task has not been yet accomplished for
the SD equation governing the dynamics of PT Green's functions~
\cite{Corn88},
we will consider the analytic continuation of our results
as a plausible assumption. We will therefore
carry out our
diagrammatic approach in terms of Feynman graphs and then continue
analytically our results to describe the physics
of unstable particles.

\setcounter{equation}{0}
\section{Issues of resummation in the PT}
\indent

Even though the PT has been developed in detail to one-loop, its
generalization to higher orders
has not yet been presented in the literature.
In this section we
will briefly outline how this generalization proceeds;
the full
presentation will be given elsewhere \cite{JPPrep}.

Here we will  focus particularly
on issues of resummation, and show that the gauge-invariant
PT self-energy may be resummed in the same way as one carries out the
Dyson summation for the conventional self-energy.
In other words,
the PT self-energies have {\em the same} resummation
properties as regular self-energies.
The crucial point is that,
even though contributions from vertices and boxes are instrumental
for the definition of the PT self-energies,
their resummation
does {\em not} require a corresponding resummation
of vertex
or box parts. In order to see that,
consider the usual Dyson series for the conventional self-energy
of QCD.
The building blocks of this series are strings of
the basic self-energy $\Pi_{\mu\nu}(q)= t_{\mu\nu}(q)\Pi(q^{2})$,
computed to a given order in perturbation theory,
which repeats itself. The net effect of the
resummation of all such strings is to bring the
quantity $\Pi(q^{2})$ in the denominator of the
free gluon propagator $\Delta_{0\, \mu\nu}$.

Let us now see how one can resum, {\em i.e.}\ bring in the
denominator the one-loop PT self-energy. To that end, consider
a string of regular one-loop self-energies (in any gauge) in QCD.
Clearly, in order to convert the
string of self-energies into a string of PT self-energies
one needs to furnish the missing pinch parts (in the same gauge).
At one loop any pinch
contribution has
the general form $[\Delta^{\mu\rho}_0(q)]^{-1}V^{P}(q)$
(for propagator-like pinch parts coming from vertices) and
$[\Delta^{\mu\rho}_0(q)]^{-1}B^{P}(q)
[\Delta^{\rho\nu}_0(q)]^{-1}$
for propagator-like pinch parts coming from boxes).
To simplify the picture
(without loss of generality) let us work in the Feynman gauge
$\xi=1$. Then at one-loop
the
only pinch contribution comes from vertices
(beyond one loop we have propagator-like pinch parts from boxes,
even for $\xi=1$).
So for each
conventional $\Pi_{\mu\nu}(q)$ we need to supply a factor
$[\Delta^{\mu\nu}_0(q)]^{-1}
\frac{1}{2}V^P (q)+ \frac{1}{2}V^P (q)[\Delta^{\mu\nu}_0(q)]^{-1}$.
Some of the necessary pinch contributions will be provided by
graphs containing at least one vertex, such as in Fig.
2(b), 2(c), and 2(d).
These existing pinch parts are however not sufficient for
converting all $\Pi_{\mu\nu}$ into $\widehat{\Pi}_{\mu\nu}$.
If we add by hand (and subsequently subtract) the missing
pieces to each $\Pi_{\mu\nu}$

\begin{itemize}

\item[(a)] The string has been converted into a string with
$\Pi_{\mu\nu}\to \widehat{\Pi}_{\mu\nu}$

\item[(b)] The left-overs, due to the presence of the inverse
$[\Delta^{\mu\nu}_0]^{-1}$ are
effectively one-particle {\em irreducible}.

\end{itemize}

To see that in detail, let us turn to the specific example
shown in Fig.~2.
The original string $L$ with two one-loop self-energies
reads (there is an overall factor $t_{\mu\nu}$ which is factored
out)
\begin{equation}
L\ =\ \frac{1}{q^2}
\left[ \Pi_1\, \left( \frac{1}{q^2} \right)\, \Pi_1\right]\,
\frac{1}{q^2} \label{LL}
\end{equation}
and is accompanied by the three strings $L_{1}$,
$L_{2}$ and $L_{3}$ shown in Figs.~2(b), 2(c), and 2(d), respectively.
After extracting the pinch contributions
from the one-loop vertices of $L_{1}$
$L_{2}$ and $L_{3}$ as is depicted in Figs.~2(e), 2(f), and 2(g), we
receive the following propagator-like contributions:
\begin{eqnarray}
L_{1}^{P}&=& \frac{1}{q^{2}}\,
\left[\, q^{2}\frac{1}{2}V^{P}_1\, \left( \frac{1}{q^{2}}\right)\,
\Pi_1\, \right]\, \frac{1}{q^{2}}\nonumber\\
L_{2}^{P}&=& \frac{1}{q^{2}}\, \left[\, \Pi_1\,
\left( \frac{1}{q^{2}}\right)\,
\frac{1}{2}V^{P}_1q^{2}\, \right]\,
\frac{1}{q^{2}}\nonumber\\
L_{3}^{P}&=&
\frac{1}{q^{2}}\, \left[\, q^{2}\frac{1}{2}V^P_1\,
\left(\frac{1}{q^{2}}\right)\,
\frac{1}{2}V^P_1 q^{2}\, \right]\, \frac{1}{q^{2}}
\label{L123}
\end{eqnarray}
Returning to $L$, we know that in order for
a $\Pi$ to be converted into a $\widehat{\Pi}$
an amount $(q^{2}\frac{1}{2}V^{P}+\frac{1}{2}V^{P}q^{2})$ must be added.
Let us call $\widehat{L}$
the corresponding string containing two $\widehat{\Pi}_1$
instead of two $\Pi$. Let us see how we can construct it from
the existing pieces:
\begin{eqnarray}
\widehat{L} &=&\frac{1}{q^{2}}\,
\left[\, \widehat{\Pi}_1\, \left(\frac{1}{q^{2}}\right)\,
\widehat{\Pi}_1\, \right]\,
\frac{1}{q^{2}}\nonumber\\
&=&\frac{1}{q^{2}}\, \left[\, \Pi_1+q^{2}\frac{1}{2}V^{P}_1
+\frac{1}{2}V^{P}_1q^{2}\, \right]\, \left(\frac{1}{q^{2}}\right)\,
\left[\, \Pi_1+q^{2}\frac{1}{2}V^{P}_1+\frac{1}{2}V^{P}_1 q^{2}\, \right]\,
\frac{1}{q^{2}}\nonumber\\
&=& L+L_{1}^{P}+L_{2}^{P}+L_{3}^{P}+\frac{1}{q^{2}}R\frac{1}{q^{2}}
\label{Lhat}
\end{eqnarray}
where
\begin{equation}
R= \Pi_1 \frac{1}{2}V^{P}_1+ \frac{1}{2}V^{P}_1\Pi_1+
\frac{1}{4}(V^{P}_1V^{P}_1 q^{2}+q^{2}V^{P}_1V^{P}_1+V^{P}_1 q^{2}V^{P}_1)
\label{R}
\end{equation}
We see that in addition to the existing pieces
$L$, $L_{1}^{P}$, $L_{2}^{P}$, and $L_{3}^{P}$,  one needs to
supply $R$. As advertised, $R$ has the very important property
that it is
effectively one-particle irreducible.
So, $R$ has the same structure as the
one-particle irreducible two-loop self-energy graphs shown in Fig.~3.
Evidently, $-R$ together with
the genuine two-loop vertex and box pinch contributions displayed in
Fig.~4 will then convert the conventional two-loop self-energy into the
g.i.\ two-loop PT self-energy.
So, the general form of the QCD propagator-like pinch
contributions in the Feynman gauge,
to a given loop order $n$ in perturbation theory,
has the form $t_{\mu\nu}(q)\Pi^{P}_{n}(q^{2})$,
with
\begin{equation}
\Pi^{P}_{n}(q^{2})= q^{2}V_{n}^{P}(q^{2})+
{(q^{2})}^{2}B_{n}^{P}(q^{2})+R_{n}^{P}(q^{2})~.
\label{GFPT}
\end{equation}
For example, propagator-like pinch contributions from one-loop
vertex graphs have the general form of the first term in the r.h.s
of Eq.~(\ref{GFPT}), whereas one-loop contributions from boxes have the
general form of the second term.
The $R_{n}(q^{2})$ contains
contributions of all terms described in (b). Clearly,
$R_{1}(q^{2})=0$, but
$R_{n}(q^{2})\ne 0$ for $n>1$.
For example, for $n=2$ we have that $R_{2}^{P}$ is the negative
of $R$ of Eq.~(\ref{R}). In this notation, $R^P_2$ reads
\begin{equation}
R_{2}^{P}(q^{2})\ =\ -R\ =\
- \left( \Pi_1 V^{P}_1+ \frac{3}{4}q^{2}V^{P}_1V^{P}_1 \right)~.
\label{R2}
\end{equation}
Obviously, the $R_{n}^{P}$ terms consist in general
of products
of lower order
conventional self-energies $\Pi_{k}(q^{2})$,
and lower order pinch contributions $V^{P}_{\ell}$ and/or
$B^{P}_{\ell}$, with $k+\ell=n$.

We emphasize that the procedure described above has
not been tailored for the particular needs of the present problem, but
it is of general validity. In fact, this is the way how the PT must be
generalized to higher orders: one has to first convert subset of
diagrams {\em locally} into the corresponding PT subsets
using the results of the previous order, by
adding (and subsequently subtracting)
the appropriate pinch parts, every time they are not
present.
Due to their characteristic
structure the extra pieces give rise to diagrams which then
can (and they should) be allotted to the remaining graphs, and they are
crucial for their gauge independence.
In this way, one can rewrite
the $S$ matrix at each order in perturbation theory, into
manifestly g.i.\ sub-amplitudes, with the characteristic
properties one knows from one loop. In fact, it is of particular
importance to explicitly demonstrate that the procedure
described above will indeed give rise to a g.i.\
two-loop self-energy, whose divergent part will coincide with
the g.i.\ two-loop QCD $\beta$ function.
Results in this direction will be presented in detail in
\cite{JPPrep}.

We conclude this section with some technical remarks.
It has been known for years that when computing the PT Green's
functions any convenient gauge may be chosen, as long as one
properly accounts for the pinch contributions within that gauge
\cite{JMC}.
In the context of the ``renormalizable'' $R_{\xi}$ gauges the most
convenient gauge-fixing choice is the Feynman gauge ($\xi=1$).
This is so because the longitudinal parts of the gauge boson
propagators, which can pinch, vanish for $\xi=1$, and the
only possibility for pinching stems from the tree-boson vertices.
As was recently realized \cite{BFM}, the task of the PT re-arrangement of the
$S$ matrix can be further facilitated, if one quantizes the theory
in the context of the Background Field Method (BFM) \cite{Abbott}.
Even though the Feynman rules obtained via the BFM
are rather involved, they
become particularly convenient
for one-loop pinching, if one chooses
the Feynman gauge ($\xi_{Q}=1$)
inside the quantum loops.
In fact, all possible one-loop pinch contribution
are zero in this gauge, {\em e.g.}\
$V_{1}^{P}|_{\xi_{Q}=1}=B_{1}^{P}|_{\xi_{Q}=1}=0$.
Consequently, the one-loop PT
Green's functions (which one can obtain for {\it every}
gauge) are {\em identical}
to the {\em conventional} Green's functions,
calculated in the Feynman gauge of the BFM.
This correspondence between PT and BFM at $\xi_{Q}=1$ breaks down
for the two-loop purely bosonic part
\cite{JMCPC}.
Therefore, $V_{n}^{P}|_{\xi_{Q}=1}\neq 0$ and
$B_{n}^{P}|_{\xi_{Q}=1} \neq 0$, for $n>1$.
The technical details
leading to these conclusions will be presented in \cite{JPPrep}.

\setcounter{equation}{0}
\section{PT resummation with non-conserved currents}
\indent

We now describe how to generalize the form of $\widehat{T}_1$,
presented in Eq~(\ref{qwer})
for the one-loop case, to higher orders.
In particular we want to show that
when the external currents are non-conserved, all possible
g.i.\ propagator-like strings assume the form of
Eq.~(\ref{T1L1}). For definiteness, we concentrate on the case
where the external currents are charged. Exactly analogous
arguments hold for neutral currents.
To accomplish that we must follow a three-step procedure:

\begin{itemize}
\item[(a)] As described in the previous section,
if we work at loop order $n$ in perturbation theory, the
strings containing
conventional
$\Pi_{\mu\nu}$, $\Theta_{\mu}$ and
$\Omega$ self-energies
(of individual order less that $n$, but of
combined order $n$)
must be converted to the corresponding
PT strings containing ${\widehat{\Pi}}_{\mu\nu}$,
${\widehat{\Theta}}_{\mu}$, and $\widehat{\Omega}$,
{\em i.e.}\ we must replace conventional with ``hatted'' quantities.
In doing so we use the formulas and methodology developed
in \cite{JPself}.
As in the previous section, we assume that
the necessary pinch parts form the lower orders
are known; in particular,
the missing pinch contributions are supplied by hand,
and subsequently subtracted.
The left-overs are effectively one-particle irreducible
and will be added to the corresponding
$\Pi_{\mu\nu}$, $\Theta_{\mu}$ and $\Omega$ of order $n$.
All such terms, together with the normal pinch parts
from box and vertex graphs of order $n$,
will finally give rise to the ${\widehat{\Pi}}_{\mu\nu}$
${\widehat{\Theta}}_{\mu}$, and $\widehat{\Omega}$ of that order.

\item[(b)] By close analogy to Eq.~(\ref{GFPT}), the
general form of the transverse propagator-like pinch
contribution to the massive gauge boson is given by
\begin{equation}
\Pi^{P}_{n}(q^{2})= (q^{2}-m_{0}^{2})V_{n}^{P}(q^{2})+
{(q^{2}-m_{0}^{2})}^{2}B_{n}^{P}(q^{2})+R_{n}^{P}(q^{2})~.
\label{GFPT2}
\end{equation}
The generic form of $R_{n}^{P}$ is also very similar; the
$R_{2}$ for example is simply
\begin{equation}
R_2\ =\
-\left( \Pi_1 V_1^{P} + \frac{3}{4}(q^{2}-m_{0}^{2})V_1^{P}V_1^{P}\right)
\label{R22}
\end{equation}
Of course, the closed expressions of
the individual $V_{n}^{P}$, $B_{n}^{P}$,
and $R_{n}^{P}$ are in general different from the QCD case.
It is important to notice that $R_{n}$ contains
a non-zero number of terms which are
{\em not} explicitly proportional to $(q^{2}-m_{0}^{2})$; this
is so because, as explained above,
the explicit ${[\Delta^{\mu\nu}_{0}]}^{-1}$ in front of the
$\Pi_{k}(q^{2})$ cancels against one of the $\Delta^{\mu\nu}_{0}$
of the string.

\item[(c)] When all possible strings have been converted to PT
strings, one can show
that due to the Ward identities in Eqs.~(\ref{PT1})--(\ref{PT3}), they
finally reorganize themselves
into two different types of g.i.\
strings, ${\widehat{T}}_{1}^{t}$ and ${\widehat{T}}_{1}^{\ell}$
of the form
\begin{equation}
{[{\widehat{T}}_{1}^{t}]}_{\mu\nu}= t_{\mu\nu}D_{0}
{\widehat{\Pi}}_{T}^{i_{1}}
D_{0}{\widehat{\Pi}}_{T}^{i_{2}}
D_{0}\{...\}D_{0}
{\widehat{\Pi}}_{T}^{i_{k-1}}
D_{0}
{\widehat{\Pi}}_{T}^{i_{k}}D_{0}
\label{String1}
\end{equation}
and
\begin{equation}
{[{\widehat{T}}_{1}^{\ell}]}_{\mu\nu}=
\ell_{\mu\nu}[\frac{1}{M^{2}}]
{\widehat{\Pi}}_{L}^{i_{1}}[\frac{1}{M^{2}}]{\widehat{\Pi}}_{L}^{i_{2}}
[\frac{1}{M^{2}}]\{...\}[\frac{1}{M^{2}}]
{\widehat{\Pi}}_{L}^{i_{k-1}}[\frac{1}{M^{2}}]{\widehat{\Pi}}_{L}^{i_{k}}
[\frac{1}{M^{2}}]\, .
\label{String2}
\end{equation}
Here, $D_{0}\equiv D_{0}^{(\xi=1)}=(q^2 - M^2 )^{-1}$
defined in Eq.\ (\ref{D0}),
${\widehat{\Pi}}_{T}^{i_{j}}$
is the PT transverse $WW$ self-energy of loop order
$i_{j}$, ${\widehat{\Pi}}_{L}^{i_{j}}$ is the PT $G^{-}G^{-}$
self-energy, and $\sum_{j=1}^{k}(i_{j})=n$.
Of course, for resummation purposes to a given
loop order $n$, we have to identify all the possible
combinatorial strings of self-energies in Eqs.\ (\ref{String1})
and (\ref{String2}), which will yield the
resummed propagator of order $n$.

\end{itemize}

To give a concrete example, let us consider
the entire set of possible strings
at $n=2$, for the process $e^-\bar{\nu}_e\to \mu^-\bar{\nu}_\mu$
shown in Fig.~5. Their explicit expressions are:
\begin{eqnarray}
(\mbox{a})&=& [U_{\mu\rho}-\frac{q_{\mu}q_{\rho}}{M^{2}}D_{0}^{(\xi)}]
{\widehat{\Pi}}^{\rho\sigma}
[U_{\sigma\lambda}-\frac{q_{\sigma}q_{\lambda}}{M^{2}}D_{0}^{(\xi)}]
{\widehat{\Pi}}^{\lambda\tau}
[U_{\tau\nu}-\frac{q_{\tau}q_{\nu}}{M^{2}}D_{0}^{(\xi)}]
\nonumber\\
(\mbox{b})&=& \frac{q_{\mu}}{M}D_{0}^{(\xi)}{\widehat{\Theta}}^{\rho}
[U_{\rho\sigma}-\frac{q_{\rho}q_{\sigma}}{M^{2}}D_{0}^{(\xi)}]
{\widehat{\Pi}}^{\sigma\tau}
[U_{\tau\nu}-\frac{q_{\tau}q_{\nu}}{M^{2}}D_{0}^{(\xi)}]
\nonumber\\
(\mbox{c})&=& [U_{\mu\rho}-\frac{q_{\mu}q_{\rho}}{M^{2}}D_{0}^{(\xi)}]
{\widehat{\Theta}}^{\rho}D_{0}^{(\xi)}{\widehat{\Theta}}^{\tau}
[U_{\tau\nu}-\frac{q_{\tau}q_{\nu}}{M^{2}}D_{0}^{(\xi)}]
\nonumber\\
(\mbox{d})&=& \frac{q_{\mu}}{M}D^{(\xi)}{\widehat{\Omega}}
D_{0}^{(\xi)}{\widehat{\Theta}}^{\tau}
[U_{\tau\nu}-\frac{q_{\tau}q_{\nu}}{M^{2}}D_{0}^{(\xi)}]
\nonumber\\
(\mbox{e})&=& [U_{\mu\rho}-\frac{q_{\mu}q_{\rho}}{M^{2}}D_{0}^{(\xi)}]
{\widehat{\Theta}}^{\rho}D^{(\xi)}{\widehat{\Omega}}
D_{0}^{(\xi)}\frac{q_{\nu}}{M}\nonumber\\
(\mbox{f})&=& \frac{q_{\mu}}{M}D_{0}^{(\xi)}{\widehat{\Omega}}D_{0}^{(\xi)}
{\widehat{\Omega}}D_{0}^{(\xi)}\frac{q_{\nu}}{M}
\nonumber\\
(\mbox{g})&=& [U_{\mu\rho}-\frac{q_{\mu}q_{\rho}}{M^{2}}D_{0}^{(\xi)}]
{\widehat{\Pi}}^{\rho\sigma}
[U_{\sigma\tau}-\frac{q_{\sigma}q_{\tau}}{M^{2}}D_{0}^{(\xi)}]
{\widehat{\Theta}}^{\tau}D_{0}^{(\xi)}\frac{q_{\nu}}{M}
\nonumber\\
(\mbox{h})&=& \frac{q_{\mu}}{M}D_{0}^{(\xi)}{\widehat{\Theta}}^{\rho}
[U_{\rho\sigma}-\frac{q_{\rho}q_{\sigma}}{M^{2}}D_{0}^{(\xi)}]
{\widehat{\Theta}}^{\sigma}D_{0}^{(\xi)}\frac{q_{\nu}}{M}
\end{eqnarray}
 It is now straightforward to prove that due to the Ward identities
of Eqs.\ (\ref{PT2}) and (\ref{PT3})
 all remaining $\xi$-dependences cancel.
To see that we can simply isolate powers of $D^{(\xi)}$
and verify that their cofactors,
by virtue of the Ward identities add up to zero
(this is essentially the approach presented in \cite{JPKP2}).
Equivalently, we notice that the above strings may be
combined pairwise [(a) with (b), (c) with (d), (e) with (f),
and (g) with (h)], to yield, (after using Eqs.\ (\ref{PT2}) and
(\ref{PT3})):
\begin{eqnarray}
(\mbox{a})+(\mbox{b})&=& U_{\mu\rho}
{\widehat{\Pi}}^{\rho\sigma}
[U_{\sigma\lambda}-\frac{q_{\sigma}q_{\lambda}}{M^{2}}D_{0}^{(\xi)}]
{\widehat{\Pi}}^{\lambda\tau}
[U_{\tau\nu}-\frac{q_{\tau}q_{\nu}}{M^{2}}D_{0}^{(\xi)}]
\nonumber\\
(\mbox{c})+(\mbox{d})&=& U_{\mu\rho}
{\widehat{\Theta}}^{\rho}D_{0}^{(\xi)}{\widehat{\Theta}}^{\tau}
[U_{\tau\nu}-\frac{q_{\tau}q_{\nu}}{M^{2}}D_{0}^{(\xi)}]
\nonumber\\
(\mbox{e})+(\mbox{f})&=& U_{\mu\rho}
{\widehat{\Theta}}^{\rho}D^{(\xi)}{\widehat{\Omega}}
D_{0}^{(\xi)}\frac{q_{\nu}}{M}
\nonumber\\
(\mbox{g})+(\mbox{h})&=& U_{\mu\rho}
{\widehat{\Pi}}^{\rho\sigma}
[U_{\sigma\tau}-\frac{q_{\sigma}q_{\tau}}{M^{2}}D_{0}^{(\xi)}]
{\widehat{\Theta}}^{\tau}D_{0}^{(\xi)}\frac{q_{\nu}}{M}
\end{eqnarray}
We can then further combine (a)+(b) with (g)+(h) and (c)+(d) with (e)+(f):
\begin{eqnarray}
(\mbox{a})+(\mbox{b})+(\mbox{g})+(\mbox{h})
&=& U_{\mu\rho}{\widehat{\Pi}}^{\rho\sigma}
[U_{\sigma\lambda}-\frac{q_{\sigma}q_{\lambda}}{M^{2}}D_{0}^{(\xi)}]
{\widehat{\Pi}}^{\lambda\tau}U_{\tau\nu}
\nonumber\\
(\mbox{c})+(\mbox{d})+(\mbox{e})+(\mbox{f})
&=& U_{\mu\rho}{\widehat{\Theta}}^{\rho}
D_{0}^{(\xi)}{\widehat{\Theta}}^{\tau}U_{\tau\nu}
\end{eqnarray}
which finally gives
\begin{equation}
{[{\widehat{T}}_{1}]}_{\mu\nu}=
U_{\mu\rho}{\widehat{\Pi}}^{\rho\sigma}U_{\sigma\lambda}
{\widehat{\Pi}}^{\lambda\tau}U_{\tau\nu}
\label{FinalT}
\end{equation}
We may now write the
${[{\widehat{T}}_{1}]}_{\mu\nu}$ of Eq.\ (\ref{FinalT}) as the
sum of two pieces, ${[{\widehat{T}}_{1}^{t}]}_{\mu\nu}$
 and ${[{\widehat{T}}_{1}^{\ell}]}_{\mu\nu}$, of the general
form advertised in Eqs.\ (\ref{String1})
and (\ref{String2}), respectively.
Indeed, using the identity of Eq.\ (\ref{U1}),
and the Ward identities, we obtain
\begin{eqnarray}
{[{\widehat{T}}_{1}]}_{\mu\nu}&=&
t_{\mu\nu}D_{0}{\widehat{\Pi}}_{T}D_{0}{\widehat{\Pi}}_{T}D_{0}
+ \ell_{\mu\nu}[\frac{1}{M^{2}}]
{\widehat{\Pi}}_{L}[\frac{1}{M^{2}}]
{\widehat{\Pi}}_{L}[\frac{1}{M^{2}}]\nonumber\\
&=& {[{\widehat{T}}_{1}^{t}]}_{\mu\nu}+{[{\widehat{T}}_{1}^{\ell}]}_{\mu\nu}
\end{eqnarray}
It is obvious how to generalize the above arguments
to an arbitrary loop order $n$, which will formally lead to the resummed
propagator, $\hat{\Delta}_{\mu\nu}$, stated in Eq.\ (\ref{Dres})
in the limit $n\to \infty$.

\setcounter{equation}{0}
\section{ The position of the pole in the PT}
\indent

Another important issue in the context of the PT is the following.
It is known that even though the conventional gauge boson
self-energy is gauge dependent, the position of the pole is
a g.i.\ quantity~\cite{WV,RGS}. On the other hand, the PT
self-energy is by construction g.i.\ for all values
of $q^{2}$, and therefore its pole is also guaranteed to be
g.i. Given the fact that the pole position
of the conventional propagator is related to physical
quantities (mass and width) it is important to inquire,
whether or not the PT pole position is different from that of
the conventional one. It turns out that, to any order
in perturbation theory the two poles are {\em identical}.
Put in different words, if one works at loop order
$n$ in perturbation theory, the two poles differ by a
{\em gauge independent} amount,
which is of order $n+1$.
This fact may come as no surprise
since the PT seems to have the general
property of not affecting quantities which are already
g.i.

In order to gain some intuition, let us first concentrate on the
simpler case of a stable particle, and show that its mass does not
get shifted by the PT.
The conventional propagator $\Delta_{\mu\nu}(q)$
(computed at some gauge), and the PT propagator
${\hat{\Delta}}_{\mu\nu}(q)$
have the form:
\begin{equation}
\Delta_{\mu\nu}(q)=\frac{-ig_{\mu\nu}}{q^{2}-m_{0}^{2}-\Pi(q^{2})}+ \cdots
\end{equation}
and
\begin{equation}
{\hat{\Delta}}_{\mu\nu}(q)=
\frac{-ig_{\mu\nu}}{q^{2}-m_{0}^{2}-\widehat{\Pi}(q^{2})}+ \cdots,
\end{equation}
where the ellipses denote the omission of terms proportional
to $q^{\mu}q^{\nu}$.
The corresponding masses $m$ and $\hat{m}$, respectively,
are defined as the solution of the following two equations
\begin{equation}
m^{2}=m_{0}^{2}+\Pi(m^{2})
\label{SP1}
\end{equation}
and
\begin{equation}
{\hat{m}}^{2}=m_{0}^{2}+\widehat{\Pi}({\hat{m}}^{2})
\label{SP2}
\end{equation}
In perturbation theory clearly
$m^{2}=m_{0}^{2}+\sum_{1}^{\infty}g^{2n}C_{n}$
and
${\hat{m}}^{2}=m_{0}^{2}+\sum_{1}^{\infty}g^{2n}{\widehat{C}}_{n}$,
and to zeroth order $m^{2}={\hat{m}}^{2}=m_{0}^{2}$.
Therefore
\begin{equation}
{\hat{m}}^{2}-m_{0}^{2}=O(g^{2})
\label{SP3}
\end{equation}
At one loop it is easy to see what happens.
To begin with, to any order in perturbation theory
\begin{equation}
\widehat{\Pi}_{n}(q^{2})= \Pi_{n}(q^{2})+\Pi_{n}^{P}(q^{2})~.
\end{equation}
The general form of the one-loop $\Pi_{1}^{P}(q^{2})$, in any gauge,
is given by
\begin{equation}
\Pi_{1}^{P}(q^{2})= (q^{2}-m_{0}^{2})V^{P}_{1}(q^{2})+
{(q^{2}-m_{0}^{2})}^{2}B^{P}_{1}(q^{2})~,
\label{SP4}
\end{equation}
and of course $R_{1}^{P}=0$ for every gauge;
in addition, in the Feynman gauge $B^{P}_{1}=0$
So, from Eqs.\ (\ref{SP1})--(\ref{SP4})
and assuming that $V^{P}_{1}({\hat{m}}^{2})$ and
$B^{P}_{1}({\hat{m}}^{2})$
are non-singular,
we have that
\begin{eqnarray}
{\hat{m}}_{1}^{2}&=m_{0}^{2}+\Pi_{1}({\hat{m}}^{2})+ O(g^{4})\\
&={m}_{1}^{2}+O(g^{4})
\end{eqnarray}
from which follows that $C_{1}=\widehat{C}_{1}$.

The non-trivial step in generalizing this proof to higher orders
is to observe that
not all pinch contributions
in the previous equation contribute terms of higher
order.
Indeed, as already mentioned in Section 4,
the $R^{P}$ terms of Eq.~(\ref{GFPT2})
do not always have a characteristic factor $(q^{2}-m_{0}^{2})$
in front, because it has been
cancelled by an internal propagator of the string.
Such terms are {\em not} of higher order, as is the
case with the graphs which are of the form given in
Eq.\ (\ref{SP4}).
To see why such contributions are instrumental
for our proof, let us repeat the
previous calculation, in
the two-loop case. At the two-loop order,
$m^{2}$ and ${\hat{m}}^{2}$ are given by:
\begin{equation}
m^{2}=m_{0}^{2}+\Pi_{1}(m^{2})+\Pi_{2}(m^{2})
\end{equation}
and
\begin{equation}
{\hat{m}}^{2}=m_{0}^{2}+\Pi_{1}({\hat{m}}^{2})+
\Pi_{2}({\hat{m}}^{2})+ \Pi_{1}^{P}+ \Pi_{2}^{P}
\end{equation}
where
\begin{eqnarray}
\Pi_{1}^{P}({\hat{m}}^{2})+\Pi_{2}^{P}({\hat{m}}^{2})
&=& ({\hat{m}}^{2}-m_{0}^{2})
[ V_{1}^{P}({\hat{m}}^{2})+V_{2}^{P}({\hat{m}}^{2}) ] +
{({\hat{m}}^{2}-m_{0}^{2})}^{2}
[ B_{1}^{P}({\hat{m}}^{2})+B_{2}^{P}({\hat{m}}^{2})]\nonumber\\
&&+ R_{2}^{P}({\hat{m}}^{2})\, .
\label{SP6}
\end{eqnarray}
We want to show that
$\Pi_{1}^{P}({\hat{m}}^{2})+\Pi_{2}^{P}({\hat{m}}^{2})=O(g^{6})$;
substituting~ ${\hat{m}}^{2}-m_{0}^{2}=\Pi_{1}({\hat{m}}^{2})+ O(g^{4})$
into Eq.~(\ref{SP6}), and neglecting terms of
$O(g^{6})$ or higher,
 we find
\begin{equation}
\Pi_1^{P}({\hat{m}}^{2})+\Pi_2^{P}({\hat{m}}^{2})=
R_{2}^{P}({\hat{m}}^{2})+
\Pi_{1}({\hat{m}}^{2})V^P_{1}({\hat{m}}^{2})+
O(g^{6})=0+O(g^{6})
\end{equation}
In the final step we have used Eq.~(\ref{R22}) at
$q^{2}={\hat{m}}^{2}$, {\em i.e.}
\begin{eqnarray}
R_{2}^{P}({\hat{m}}^{2})&=&
-\Pi_1({\hat{m}}^{2})
V^{P}_1({\hat{m}}^{2}) -\frac{3}{4}(m^{2}-m_{0}^{2})
V^{P}_1({\hat{m}}^{2})V^{P}_1({\hat{m}}^{2})\nonumber\\
&=&-\Pi_1({\hat{m}}^{2})V_1^{P}({\hat{m}}^{2})+O(g^{6})
\label{XX}
\end{eqnarray}

The generalization of the previous proof to an arbitrary
loop order $n$ in perturbation theory proceeds by induction.
First of all, to simplify things we will work in the Feynman gauge.
In that case, the general form of the $R^{P}$ terms becomes
\begin{equation}
R^{P}_{n}= (q^{2}-m_{0}^{2})v^{P}_{n}+ {(q^{2}-m_{0}^{2})}^{2}b^{P}_{n}+
{\tilde R}^{P}_{n}\, ,
\end{equation}
where ${\tilde R}^{P}_{n}$ is the part of
$R^{P}_{n}$ which is of $O(g^{2n})$ at $q^{2}=m^{2}_{0}$,
whereas the rest is $O(g^{2(n+1)})$. For example, from $R^{P}_{2}$
of Eq.~(\ref{R22}), or equivalently Eq.~(\ref{XX}), we have that
${\tilde R}^{P}_{2}(q)=-\Pi_1(q)V_1^{P}(q)$.
Finally, we define ${\cal V}^{P}_{n}$ and ${\cal B}^{P}_{n}$ as
follows:
\begin{eqnarray}
{\cal V}^{P}_{n}&=& V^{P}_{n}+v^{P}_{n}\, , \nonumber\\
{\cal B}^{P}_{n}&=& B^{P}_{n}+b^{P}_{n}\, .
\end{eqnarray}
Let us now assume that ${\hat{m}}^{2}=m^{2}$, up to order $n-1$, {\em i.e.}\
$C_{k}={\widehat{C}}_{k}$, for every $k\le n-1$.
The expression for ${\hat{m}}^{2}$ to order $n$ is
\begin{equation}
{\hat{m}}^{2}\ =\ m_{0}^{2}\ +\ \sum_{k=1}^{n} \Pi_{k}\ +\
({\hat{m}}^{2}-m_{0}^{2})\sum_{k=1}^{n}{\cal V}^{P}_{k}\
+\ ({\hat{m}}^{2}-m_{0}^{2})^{2}
\sum_{k=1}^{n}{\cal B}^{P}_{k}\ +\ \sum_{k=1}^{n}{\tilde R}^{P}_{k}\, .
\label{SP7}
\end{equation}
Using the fact that
${\hat{m}}^{2}-m_{0}^{2}=\sum_{1}^{n-1} \Pi_{k}
+O(g^{2n})$
(from the previous order), and that,
as before, both
$({\hat{m}}^{2}-m_{0}^{2}){\cal V}_{n}^{P}$
and ${({\hat{m}}^{2}-m_{0}^{2})}^{2}{\cal B}_{n}^{P}$
are of $O(g^{2n+2})$ and higher, Eq.~(\ref{SP7}) becomes
\begin{eqnarray}
{\hat{m}}^{2}&=&m_{0}^{2}\ +\ \sum_{k=1}^{n} \Pi_{k}\ +\
\sum_{k=1}^{n-1} \Pi_{k}
\sum_{m=1}^{n-1} {\cal V}^{P}_{m}\
+\ \Big[\, \sum_{k=1}^{n-1} \Pi_{k}\Big]^2\,
\sum_{m=1}^{n-1}{\cal B}^{P}_m\ +\ \sum_{k=1}^{n}
{\tilde R}^{P}_{k}\nonumber\\
&=& m^{2}\ +\ \sum_{k=1}^{n}\sum_{\ell=1}^{k}\Pi_{\ell}
{\cal V}^{P}_{k-\ell}\ +\
\sum_{k=1}^{n}\sum_{j=1}^{k}
\sum_{\ell=1}^{j}\Pi_{\ell}\Pi_{j-\ell}{\cal B}^{P}_{k-j}\
+\ \sum_{k=1}^{n} {\tilde R}^{P}_{k}\nonumber\\
&=& m^{2}\ +\
\sum_{k=1}^{n}\left({\tilde R}^{P}_{k}+\sum_{\ell=1}^{k}\Pi_{\ell}
{\cal V}^{P}_{k-\ell}+
\sum_{j=1}^{k}\sum_{\ell=1}^{j}\Pi_{\ell}\Pi_{j-\ell}
{\cal B}^{P}_{k-j}\right)\, .
\label{BigOne}
\end{eqnarray}
It is a matter of careful counting to convince oneself that
each term of the
series in the r.h.s.\ of the last Eq.~(\ref{BigOne}) vanishes, {\em i.e.}
\begin{equation}
{\tilde R}^{P}_{k}\ +\ \sum_{\ell=1}^{k}\Pi_{\ell}
{\cal V}^{P}_{k-\ell}\ +\
\sum_{j=1}^{k}\sum_{\ell=1}^{j}\Pi_{\ell}\Pi_{j-\ell}
{\cal B}^{P}_{k-j}\
=\ 0\, ,
\label{RFormula}
\end{equation}
which means that to order $n$, ${\hat{m}}^{2}=m^{2}$,
or equivalently, $C_{n}={\widehat{C}}_{n}$, for every $n$.
In Appendix~A, we present a proof of Eq.~(\ref{RFormula}).
It is interesting to see that it is
precisely the left-over contributions we obtain when we
convert conventional strings into g.i.\ strings, which
enforce the equality between the conventional and PT poles.

\setcounter{equation}{0}
\section {The case of the unstable particle}
\indent

We now proceed to the case of an unstable particle;
we want to show that both the mass and the width remain
unshifted in the context of the PT. We will adopt the
definitions and methodology introduced by Sirlin~\cite{AS}.
Calling $s=q^{2}$, the
pole position $\bar{s}$ is defined as the solution of the
following equation:
\begin{equation}
\bar{s}=m_{0}^{2}+ \Pi(\bar{s})
\label{UP1}
\end{equation}
We adopt the following definition of mass $m$ and
width $\Gamma$ in terms of $\bar{s}$:
\begin{equation}
\bar{s}= m^{2}-im\Gamma
\label{UP2}
\end{equation}
Similarly, in the context of the PT
we define the pole position $\hat{s}={\hat{m}}^{2}-i\hat{m}
\widehat{\Gamma}$
as the solution of
\begin{equation}
\hat{s}=m_{0}^{2}+ \widehat{\Pi}(\hat{s})
\label{UP3}
\end{equation}
We want to show that
$\bar{s}=\hat{s}$ ---or equivalently, $m=\hat{m}$ and
$\Gamma=\widehat{\Gamma}$--- to every order in perturbation theory.
Since both $\Gamma$ and $\widehat{\Gamma}$ are of
$O(g^{2})$, at one loop we have just the
result of the previous section, {\em i.e.}\ $m=\hat{m}$, for $n=1$.
Going to the next order, we expand Eqs.~(\ref{UP1})
and (\ref{UP3}) up to terms of
$O(g^{4})$,
\begin{equation}
\bar{s}=m_0^2 + \Pi(m^2) - \Pi'(m^2) im\Gamma
\label{UP4}
\end{equation}
and
\begin{equation}
\hat{s}=m_0^2+ \widehat{\Pi}(\hat{m}^2)-
\widehat{\Pi}'(\hat{m}^2) i\hat{m}\widehat{\Gamma}
\label{UP5}
\end{equation}
where $\Pi'(m^2)\equiv d\Pi(q^2)/dq^2|_{q^2=m^2}$.
Separating real and imaginary parts (we omit the arguments $m^2$
and $\hat{m}^2$, respectively)
we have
\begin{eqnarray}
m^2 &=& m_0^2+\Re e\Pi+ m\Gamma\Im m\Pi'\, ,
\label{UP6}\\
\hat{m}^2 & =& m_0^2+
\Re e\widehat{\Pi} + \hat{m}\widehat{\Gamma} \Im m \widehat{\Pi}'\, ,
\label{UP7}
\end{eqnarray}
for the real parts, and
\begin{eqnarray}
m\Gamma &=& -\Im m\Pi +m\Gamma \Re e\Pi'\, ,
\label{UP8}\\
\hat{m}\widehat{\Gamma} &=& -\Im m\widehat{\Pi} +
\hat{m}\widehat{\Gamma} \Re e \widehat{\Pi}' \, ,
\label{UP9}
\end{eqnarray}
for the imaginary parts.
Let us write $\hat{m}^2$ and $\hat{m}\widehat{\Gamma}$ as follows:
\begin{eqnarray}
\hat{m}^2 &=& m^2 + \epsilon_1\, ,
\label{UP10}\\
\hat{m}\widehat{\Gamma}& =& m\Gamma+\epsilon_2\, ,
\label{UP11}
\end{eqnarray}
where
\begin{eqnarray}
\epsilon_1 &=&\Re e\Pi^P+
\hat{m}\widehat{\Gamma} \Im m {\Pi^P}' \, ,
\label{UP12}\\
\epsilon_2 &=& -\Im m\Pi^P+\hat{m}\widehat{\Gamma} \Re e{\Pi^P}' \, .
\label{UP13}
\end{eqnarray}
In Eqs.\ (\ref{UP12}) and (\ref{UP13}), $\Pi^P$
is the total pinch contribution to order
$g^{4}$, {\em i.e.}\ $\Pi^P=\Pi^P_1+\Pi^P_2$, with the general form
given in Eq.~(\ref{SP6}).
We now want to show that
both $\epsilon_{1}$ and $\epsilon_{2}$ are of $O(g^6)$.
Using again Eq.~(\ref{SP6}) we have that
\begin{equation}
\Re e\Pi^{P}=\Re e\Pi_{1} \Re eV_{1}^{P} + \Re e R_{2}+ O(g^{6})
\label{UP14}
\end{equation}
and
\begin{eqnarray}
\hat{m}\widehat{\Gamma} \Im m{\Pi^P}' &=&
[ \Im m V_1^{P} + O(g^4) ] [-\Im m \Pi_1 + O(g^4) ]\nonumber\\
&=& -\Im m V_1^{P} \Im m \Pi_1 + O(g^{6})
\label{UP15}
\end{eqnarray}
Therefore, up to terms of $O(g^{6})$
\begin{eqnarray}
\epsilon_1&=& \Re e R_2^{P} + \Re e V_1^{P} \Re e\Pi_1
- \Im m V_1^{P} \Im m \Pi_1
\nonumber\\
&=& \Re e (R_2^{P} + \Pi_1 V_1^{P} )\nonumber\\
&=& 0\, ,
\end{eqnarray}
where we used Eq.~(\ref{XX}).
Similarly, using the fact that to $O(g^{4})$
\begin{eqnarray}
\Im m\Pi^P &=&\Im m R_2^{P} + \Im m [ (\hat{m}^2-m_0^2)V_1^{P}
 +O(g^6)]\nonumber\\
&=& \Im m R_2^{P}  + \Im m [V_1^{P} \Re e \Pi_1 + O(g^6)]\nonumber\\
&=& \Im m R_2^{P} + \Re e \Pi_1 \Im m V_1^{P}
\end{eqnarray}
and
\begin{equation}
\hat{m}\widehat{\Gamma} \Re e {\Pi^P}' = -\Re e V_1^{P} \Im m \Pi_1 + O(g^6)
\end{equation}
we have
\begin{eqnarray}
\epsilon_2 &=&- \Im m R_2^{P} - \Re e\Pi_1 \Im m V_1^{P}
 - \Re e V_1^{P} \Im m\Pi_1
\nonumber\\
&=& -\Im m (R_2^{P} + \Pi_1 V_1^{P} ) \nonumber\\
&=& 0
\end{eqnarray}
where again Eq.~(\ref{R22}) was used.

It is straightforward to generalize this result to an arbitrary
order $n$ in perturbation theory. One should simply notice that
the formula of Eq.~(\ref{R22}) and its generalization to
higher orders given by Eq.~(\ref{RFormula}) is crucial to obtain
a general proof. In particular, we have seen in Section~3 that the
extension of the PT to higher orders has given rise to new PT terms,
$R_{n}^{P}$, which guarantee that the position of the pole remains
unchanged.

\setcounter{equation}{0}
\section{Unitarity and related properties}
\indent

In this section, we will analyze issues of unitarity
pertinent to a consistent $S$-matrix perturbation theory
involving
unstable particles. In particular, we will mainly focus
on the optical theorem, which is a direct consequence of the unitarity
of the $S$ matrix, and prescribes the form of the perturbative
expansion for the transition operator $T$.

The $T$-matrix element of a reaction $i\to f$
is defined via the relation
\begin{equation}
\langle f | S | i \rangle\ =\ \delta_{fi}\ +\ i(2\pi )^4
\delta^{(4)}(P_f - P_i)\langle f | T|i\rangle ,\label{tmatrix}
\end{equation}
where $P_i$ ($P_f$) is the sum of all initial (final) momenta
of the $| i\rangle$ ($| f \rangle$) state. Furthermore, imposing
the unitarity relation $S^\dagger S = 1$ leads to the optical
theorem:
\begin{equation}
\langle f|T|i\rangle - \langle i |T|f\rangle^*\ =\
i\sum_{i'} (2\pi )^4\delta^{(4)}(P_{i'} - P_i)\langle i' | T | f \rangle^*
\langle i' | T | i \rangle. \label{optical}
\end{equation}
In Eq.~(\ref{optical}), the sum $\sum_{i'}$ should be understood to be
over the whole phase space and spins of all possible on-shell
intermediate particles $i'$.
A corollary of this theorem is obtained if $i=f$. In this particular case,
we have
\begin{equation}
\Im m \langle i|T|i\rangle\ =\ \frac{1}{2}
\sum_f (2\pi )^4\delta^{(4)}(P_f - P_i) |\langle f | T|i \rangle |^2.
\label{absorptive}
\end{equation}
In the conventional $S$-matrix theory with stable
particles, Eqs.~(\ref{optical}) and (\ref{absorptive}) hold also
perturbatively. To be precise, if one expands the transition
operator in power series of the coupling constants, say $g$, as
$T=T^{(1)}+T^{(2)}+\cdots + T^{(n)} + \cdots$, in a given
order $n$ one has
\begin{equation}
T^{(n)}_{fi}-T^{(n)*}_{if}\ =\ i\sum_{i'} (2\pi )^4
\delta^{(4)}(P_{i'}-P_i) \sum\limits_{k=1}^n
T^{(k)*}_{i'f} T^{(n-k)}_{i'i}.\label{optpert}
\end{equation}
In a scalar model containing an unstable particle, Veltman
showed~\cite{Velt} that unitarity can be preserved by suitably
modifying the $S$-matrix perturbation theory, in which unstable
particles should always appear as intermediate states.
Obviously, the $S$-matrix perturbation expansion arising from
the truncation of the unstable particles as asymptotic states
should be reformulated accordingly.
A convincing example of how the PT algorithm gives rise to
amplitudes which,
in addition to being g.i.\ also respect unitarity,
is the calculation
of the magnetic dipole moment $\mu_{W}$
and the electric quadruple $Q_{W}$ for the $W$ boson
\cite{JPKP}.
Such quantities are of particular interest in view of the
upcoming experiments of the type $e^{+}e^{-}\rightarrow W^{+}W^{-}$ \cite{TGV}
that will be studied at the CERN Large Electron Positron collider
(LEP2), which is planned to operate at a centre of mass system (c.m.s.)
energy $s=200$~GeV.

In order to understand under what conditions an expansion based on resummed
propagators can respect the unitarity relation of Eq.~(\ref{absorptive}),
let us first consider the toy model of Ref.~\cite{Velt}. This model is
a superrenormalizable $\phi^3$-scalar theory,
which contains a light scalar, $\phi$, and
a heavy one, $\Phi$, having a mass $M_\Phi > 2 M_\phi$. In order to
provide a decay mode for the heavy scalar into two $\phi$'s, one introduces
the interaction term in the Lagrangian
\begin{equation}
{\cal L}_{int}\ =\ \frac{\lambda}{2} \phi^2(x)\Phi(x),
\end{equation}
where $\lambda$ is a non-zero coupling constant.
For concreteness,
we consider the reaction $\phi\phi \to \phi\phi$ at c.m.s.\ energies
$s\simeq M^2_\Phi$. This process proceeds via three graphs; one
resonant $s$-channel graph, and two nonresonant $t$ and $u$ graphs.
After performing a Dyson summation for the $s$-, $t$-, and $u$-channel
propagators, we arrive at the following expression for the transition
amplitude:
\begin{eqnarray}
T(s,t,u) &=& -\lambda^2 \Bigg( \frac{1}{s-M^2_\Phi +\Re e\Pi_\Phi(s)
+i\Im m\Pi_\Phi(s) } + \frac{1}{t-M^2_\Phi+\Pi_\Phi(t)}\nonumber\\
&& + \frac{1}{u-M^2_\Phi+\Pi_\Phi(u)}\Bigg), \label{Tstu}
\end{eqnarray}
where $\Pi_\Phi (q^2)$ is the irreducible two-point function of the
$\Phi\Phi$ self-energy at the one-loop order.
It is easy to verify
from Eq.~(\ref{Tstu}), that the amplitude $T(s,t,u)$ is endowed with
the analyticity property of crossing symmetry.
In other words,
the various processes can be obtained by appropriately interchanging
the Mandelstam variables $s$, $t$, and $u$; obviously
$T(s,t,u)=T(t,s,u)=\cdots $.
These crossing properties can be
naturally implemented,
when the resummed self-energies appearing in Eq.~(\ref{Tstu})
are momentum-dependent.
When crossing is applied in such a case,
the unphysical absorptive parts are killed by the kinematic $\theta $
functions, whereas the new physical absorptive contributions,
which emerge
after crossing, will regulate the resulting resonant channels.
This feature persists even if vertex
and box graphs are included. A
qualitatively similar behaviour is expected
in gauge theories;
since the resummed self-energy derived
from the PT depends on $q^2$, we
conclude that our PT
approach to gauge theories with unstable particles
respects the crossing symmetry.

We will now discuss the main reason which clearly advocates for
a $q^2$-dependent regulator, rather than a constant one. If we
consider the l.h.s.\ of Eq.~(\ref{absorptive}), we have
for the process $\phi\phi \to \phi\phi$
\begin{equation}
\Im m T(s,t,u)\ =\ \frac{\lambda^2\, \Im m\Pi_\Phi (s)}{
[s-M^2_\Phi+\Re e \Pi_\Phi (s)]^2 + [\Im m \Pi_\Phi (s)]^2},
\end{equation}
which is related to the amplitude squared of the resonant $s$-exchange
graph, say $T_s$. In fact, one finds that
\begin{equation}
\Im m T(s,t,u)\ =\ \frac{1}{2}\, \int d\mbox{LIPS}\, |T_s(s)|^2\, ,
\label{AbsTs}
\end{equation}
where LIPS stands for the Lorentz-invariant phase space for the
two on-shell $\phi$ particles. Eq.~(\ref{AbsTs}) is consistent
with Eq.~(\ref{optpert}) in a perturbative sense. At this
point it is important to notice that the unitarity relation
of Eq.~(\ref{AbsTs}) is {\em only} valid when the resummation
involves an $s$-dependent two-point function and width for the
unstable scalar $\Phi$. If a constant
width
for $\Phi$ had been considered instead,
unitarity would have been violated through
Eq.~(\ref{AbsTs}), when $s\neq M^2_\Phi$. It is therefore
evident that the regulator of a resummed propagator should
be $s$-dependent in this scalar theory.
The above problem is expected to appear if one
attempts to use a constant pole expansion in the context of
a gauge field theory.  Indeed,
there is no fundamental reason to believe
that one could consistently describe gauge theories using a
resummation procedure which is not well justified even for scalar
theories.
On the other hand, the reordering of Feynman graphs via the PT and
the resummation
of the momentum dependent
PT self-energies provides a g.i.\ solution to the problem at hand,
while, at the same time,
does not introduce residual unitarity-violating
terms in the resonant matrix element.

In what follows we will analyze some crucial aspects of the
PT algorithm in relation to the unitarity,
and underline the analogies between the PT
results in gauge theories
and some known facts from the $\phi^3$ scalar theory.
In the $\phi^3$ model, the transition amplitude of
Eq.~(\ref{Tstu}) exhibits a clear separation of the
dependence on the Mandelstam variables $s$, $t$ and $u$.
In this way, resummation can be applied to each channel
independently. Because of this property, $T(s,t,u)$
displays the
correct high-energy unitarity behaviour, and vanishes
as $s$, $t\to \infty$.
In gauge theories, this is generally not the case. For
example, consider the process $l^-\bar{\nu}_l\to W^- H$
shown in Fig.~6, where the charged lepton ($l$) is massive.
In the Born approximation, there exist two graphs: an $s$- and
$t$-mediated graph in the unitary gauge (see, also, Figs.~6(a) and 6(c)).
Taking the infinite limit of $s$ and $t$ for the $s$-channel graph,
one can verify that this amplitude alone does not
vanish. On the other hand, the total matrix
element tends to zero in the high-energy unitarity
limit. Evidently, the
$t$-exchange graph contains terms, which, when
properly taken into account,
conspire in such a way so as to give the correct high-energy unitarity
limit.
The PT algorithm accomplishes,
via the decomposition given in
Eq.~(\ref{TPT}), the same clear kinematic separation one knows from
the scalar theory.

The above discussion becomes more transparent if one employs
the Ward identities which relate the Feynman
graphs of Fig.~6(a) to those of Fig.~6(b), and the
diagram of Fig.~6(c) to that of Fig.~6(d).
For the process $l\nu_l\to W^- (p_-) H(p_H)$, we have in
an arbitrary $\xi$ gauge
\begin{eqnarray}
\frac{p_-^\mu}{M_W}\, T^{(\xi )}_{(a)\, \mu} & =&
T^{(\xi )}_{(b)}\ -\ \frac{g_w}{2M_W}\Lambda_0, \label{WH1}\\
\frac{p_-^\mu}{M_W}\, T^{(\xi )}_{(c)\, \mu} & =&
T^{(\xi )}_{(d)}\ +\ \frac{g_w}{2M_W}\Lambda_0. \label{WH2}
\end{eqnarray}
In the high-energy limit where $p_-\to \infty$, the polarization
vector, $\varepsilon_L^\mu (p_-)$,  of the longitudinal $W$ boson
approaches to $p_-^\mu/M_W$. In the Feynman gauge, the amplitudes
$T_{(d)}$ and $T_{(b)}$ vanish in the limit $s\to \infty$.
In this limit, it is easy to see that the remaining constant term in
Eq.~(\ref{WH1}) is responsible for the bad high-energy behaviour,
and can only be cancelled if a corresponding term coming from
Eq.~(\ref{WH2}) is added.
It turns out that, when loop corrections are considered,
this latter term is furnished by the relevant PT part
thus leading to a proper $s$-dependent propagator~\cite{JPself}.

An issue related to the discussion of unitarity is whether
the PT self-energy which regularizes the singular propagator
contains any unphysical absorptive parts. From Eq.~(\ref{optpert}),
one has to show that the propagator-like part $\widehat{T}_1$ of a
reaction should contain imaginary parts associated with
physical Landau singularities only, whereas the unphysical
poles related to Goldstone bosons and ghosts must vanish in the loop.
Although the PT algorithm produces a g.i.\ result for $\widehat{T}_1$,
there would still have been a problem if this procedure had introduced
some fixed unphysical poles.
A qualitative argument suggesting
that this is not the case, is that the PT results can be
obtained equally well by working {\it directly}
in the
unitary gauge~\cite{JPgeneral}, where only physical Landau poles are
present. We will also demonstrate this fact by an explicit
calculation of
the $\Im m \widehat{T}_1$ of the process $e\bar{\nu}_e\to \mu\bar{\nu}_\mu$
at the one-loop electroweak order. We will assume that only the
$W$ and $H$ particles can come kinematically on the mass shell,
as shown in Fig.~7. Then, the absorptive amplitude, $\Im m M$,
for the aforementioned process may be conveniently written
as (suppressing contraction over Lorentz indices)
\begin{eqnarray}
\Im m M & = & \tilde{\Delta}_{0H}(p_H)\Bigg[
T^{1(\xi)}_{(a)} \tilde{\Delta}_0^{(\xi)}(p_-) T^{2(\xi)}_{(a)}
+ T^{1(\xi)}_{(b)} \tilde{D}_0^{(\xi)}(p_-) T^{2(\xi)}_{(b)}
+T^{1(\xi)}_{(c)} \tilde{\Delta}_0^{(\xi)}(p_-) T^{2
(\xi)}_{(a)}\nonumber\\
&& + T^{1(\xi)}_{(a)} \tilde{\Delta}_0^{(\xi)}(p_-) T^{2(\xi)}_{(c)}
+ T^{1(\xi)}_{(d)} \tilde{D}_0^{(\xi)}(p_-) T^{2(\xi)}_{(b)}
+ T^{1(\xi)}_{(b)} \tilde{D}_0^{(\xi)}(p_-) T^{2(\xi)}_{(d)}\nonumber\\
&& + T^{1(\xi)}_{(c)} \tilde{\Delta}_0^{(\xi)}(p_-) T^{2(\xi)}_{(c)}
+ T^{1(\xi)}_{(d)} \tilde{D}_0^{(\xi)}(p_-) T^{2(\xi)}_{(d)}\Bigg]\, ,
\label{ImM}
\end{eqnarray}
where $T^1$ ($T^2$) denotes the electron (muon) mediated amplitude present
in Fig.~7, and the tilde acting on the tree-level propagators simply
projects out the corresponding absorptive parts,
as these are effectively
obtained after applying the Cutkosky rules. More explicitly, we have
\begin{eqnarray}
\tilde{\Delta}_{0H} (p_H) & = & 2\pi i\, \delta_+ (p^2_H-M^2_H)\\
\tilde{D}_0^{(\xi)} (p) & =& 2\pi i\, \delta_+ (p^2 - \xi M^2_W)\\
\tilde{\Delta}_{0\, \mu\nu}^{(\xi)} (p) & = & 2\pi i \left[
\left( - g_{\mu\nu}+\frac{p_\mu p_\nu}{M^2_W}\right) \delta_+ (p^2 -M^2_W)\,
-\, \frac{p_\mu p_\nu}{M^2_W} \delta_+ (p^2 -\xi M^2_W)\right]\nonumber\\
&=& \tilde{U}_{\mu\nu}(p)\ -\ \frac{p_\mu p_\nu}{M^2_W}\tilde{D}_0^{(\xi)}(p)
\, ,
\end{eqnarray}
with $\delta_+ (p^2-M^2) = \delta (p^2 - M^2)\theta (p^0) $.
After identifying the PT piece [$T^{i}_P=g_w\Lambda^{(i)}_0/2M_W$, with
$i=1 (:e),2(:\mu)$], which is obtained from Eq.~(\ref{WH2}) each time the
$p_-^\mu p_-^\nu$-dependent part of $\tilde{\Delta}^{(\xi)}_{0\mu\nu}$ gets
contracted with $T^{i (\xi)}_{(c)}$, we find that the imaginary
propagator-like part is
\begin{eqnarray}
\Im m \widehat{T}_1 & = & \tilde{\Delta}_{0H}(p_H)\Bigg\{
T^{1(\xi)}_{(a)} \tilde{\Delta}_0^{(\xi)}(p_-) T^{2(\xi)}_{(a)}
+ T^{1(\xi)}_{(b)} \tilde{D}_0^{(\xi)}(p_-) T^{2(\xi)}_{(b)}
+ (2\pi i)\, \Bigg[ T^1_{P} \frac{p_-^\nu}{M_W} T^{2(\xi)}_{(a)\, \nu }
\nonumber\\
&& + T^{1(\xi)}_{(a)\, \lambda} \frac{p_-^\lambda}{M_W} T^2_{P}
+ T^1_{P} T^2_{P}\Bigg] \, [\delta_+ (p^2_- - M^2_W) -
\delta_+ (p^2_- - \xi M^2_W)] \Bigg\} \, \nonumber\\
&=& \tilde{\Delta}_{0H}(p_H)\Bigg\{
T^{1(\infty)}_{(a)} \tilde{U}(p_-) T^{2(\infty)}_{(a)}
+(2\pi i)\, \Bigg[ T^1_{P} \frac{p_-^\nu}{M_W} T^{2(\infty)}_{(a)\, \nu }
\nonumber\\
&& + T^{1(\infty)}_{(a)\, \lambda} \frac{p_-^\lambda}{M_W} T^2_{P}
+ T^1_{P} T^2_{P}\Bigg] \, \delta_+ (p^2_- - M^2_W)\Bigg\}
\ +\ \delta\widehat{T}_1.
\label{dT1}
\end{eqnarray}
In the last step of Eq.~(\ref{dT1}), we have separated contributions
originating from the physical poles at $p^2_H=M^2_H$ and $p_-^2=M^2_W$
from those that occur at $p^2_-=\xi M^2_W$ and are included
in $\delta\widehat{T}_1$, where
\begin{eqnarray}
\delta\widehat{T}_1 & = & \tilde{\Delta}_{0H}(p_H)\tilde{D}_0^{(\xi)}(p_-)
\Bigg[ - T^{1(\xi)}_{(a)\, \lambda} \frac{p_-^\lambda p_-^\nu}{M^2_W}
T^{2(\xi)}_{(a)\, \nu}
+ T^{1(\xi)}_{(b)} T^{2(\xi)}_{(b)}
- T^1_{P} \frac{p_-^\nu}{M_W} T^{2(\xi)}_{(a)\, \nu }
\nonumber\\
&& - T^{1(\xi )}_{(a)\, \lambda} \frac{p_-^\lambda}{M_W} T^2_{P}
- T^1_{P} T^2_{P}\Bigg]\, .
\end{eqnarray}

Obviously, the imaginary parts coming from the physical Landau
singularities are manifestly g.i., whereas the term $\delta\widehat{T}_1$
not only should be g.i.\ because of the PT reordering, but
it should vanish
identically.
With the help of Eq.~(\ref{WH1}), it is a matter of simple algebra to show
that indeed $\delta\widehat{T}_1=0$.

It is therefore important to emphasize the conclusions of this section.
The PT algorithm can effectively disentangle the different kinematic
dependences on the Mandelstam variables $s$ and $t$ via the
decomposition given in Eq.~(\ref{TPT}), when radiative corrections
are considered.  Furthermore, this algorithm yields a
proper $q^2$-dependent propagator displaying the desired
unitarity behaviour
in the high-energy limit. The PT method
not only produces  g.i.\ analytic results but also gives rise to a
well-defined self-energy, in which all possible physical absorptive
parts are present, while unphysical Landau singularities originating
from ghosts and Goldstone bosons do not survive.
This latter property is particularly advantageous,
since we wish to
resum the $q^2$-dependent PT self-energy in order to unitarize the
singular resonant amplitude, and,
at the same time, avoid the presence
of unphysical residual absorptive phases, which could be generated
if a constant pole expansion had been used instead.

\setcounter{equation}{0}
\section{The process $\gamma e^-\to \mu^-\bar{\nu}_\mu \nu_e$}
\indent

We will study the process $\gamma e^- \to \mu^-\bar{\nu}_\mu \nu_e$,
in which two gauge $W$ bosons are involved. This process is of
potential interest at the LEP2. Furthermore, the collider TEVATRON at
Fermilab offers the possibility to study the scattering process
$qq'\to \gamma \mu^-\bar{\nu}_\mu$~\cite{BZ}.

In the Born approximation, the process $\gamma e^- \to
\mu^-\bar{\nu}_\mu \nu_e$ consists of three Feynman graphs
shown in Fig.~8,
with the gauge bosons in the unitary gauge. The transition
amplitude then reads
\begin{equation}
T(\gamma e^-\to \mu^-\bar{\nu}_\mu \nu_e )\ =\ \varepsilon^\mu_\gamma (q)\,
T_{0\mu}, \label{treeWWgamma}
\end{equation}
with
\begin{eqnarray}
T_{0\mu} & = & \Gamma_{0\rho} U_W^{\rho\nu}(p_-)\,
\Gamma^{\gamma W^- W^+}_{0\mu\nu\lambda}(q,p_-,p_+)\, U_W^{\lambda\sigma}(p_+ )
\Gamma_{0\sigma}\nonumber\\
&& +\ \Gamma_{0\rho} S^{(e)}_0 \Gamma_{0\mu}^\gamma U_W^{\rho\nu}(p_+)
\Gamma_{0\nu}\
+\ \Gamma_{0\rho} U_W^{\rho\nu}(p_-) \Gamma_{0\mu}^\gamma S^{(\mu )}_0
\Gamma_{0\nu}.
\label{TWWgamma}
\end{eqnarray}
In Eq.~(\ref{TWWgamma}),
$S^{(f)}_0=(\not\!\! p-m_f)^{-1}$ denotes the free $f$-fermion propagator,
$\Gamma_0^{\gamma W^- W^+}$
($\Gamma_{0\mu}^\gamma$) is the tree-level $\gamma WW$
($l^-l^+\gamma$) coupling,
and $p_-$ ($p_+$) is the momentum of the $W^-$ ($W^+$) boson flowing into
the $\gamma W^-W^+$ vertex.
The form of the
amplitude given in (\ref{treeWWgamma})
is gauge invariant, in the sense that it does not depend on the
gauge fixing procedure nor the gauge-fixing parameter chosen.
In the $R_{\xi}$ gauges,
for example, additional graphs with Goldstone bosons must be
included,
but at the end, the expression of (\ref{treeWWgamma}) will emerge again.
In addition,
since the action of the photonic momentum on the tree-level
$\gamma WW$ vertex triggers the elementary Ward identity
\begin{equation}
\frac{1}{e}\, q^\mu \Gamma^{\gamma W^-W^+}_{0\, \mu\nu\lambda}\ =\
U^{-1}_{W\nu\lambda}(p_+) - U^{-1}_{W\nu\lambda}(p_-)\, ,\label{qmuGamma}
\end{equation}
the electromagnetic gauge invariance of the
tree-level amplitude is evident,
{\em i.e.}\ $q^\mu {T}_{0\mu}=0$.
In Eq.~(\ref{qmuGamma}), $U^{-1}_{W\mu\nu}$ is
the inverse free propagator,
of the $W$ boson in the unitary gauge.
In general, the inverse free propagator of a vector boson, $V$,
including massless gauge bosons, such as photons and gluons,
may be obtained from Eq.~(\ref{U1}) in the same gauge. Its explicit
form is given by
\begin{equation}
U^{-1}_{V\mu\nu}(q)\ =\ t_{\mu\nu}(q)(q^2-M^2_V)\ +\ \ell_{\mu\nu}(q)M^2_V.
\end{equation}
However, since the $T_{0\mu}$
of (\ref{TWWgamma}) exhibits
a physical pole at $p^2_+=M^2_W$, the use of
a resummed propagator is needed. As we have discussed in Section
2, the naive form of a BW propagator for the singular amplitudes
violates $U(1)_{em}$ and $R_\xi$ gauge invariance.
On the other hand, the PT
method used to reorder the Feynman graphs, restores
both the $U(1)_{em}$ and the $R_\xi$ invariance of the amplitude,
which are present at the tree level.

To see that, let us concentrate on the part
$\widehat{T}_{1\mu}$ of the amplitude,
shown in Fig.~8, which contains the trilinear $\gamma WW$ vertex.
Applying the PT, and then resumming the PT self-energies
following a procedure exactly analogous
to the one described in Section~2, we arrive at the resonant
transition amplitude
(suppressing all the contracted Lorentz indices except of the photonic one):
\begin{equation}
\label{Twres}
\widehat{T}_{1\mu}\ =\ \Gamma_0 \hat{\Delta}_W
(\Gamma^{\gamma W^- W^+}_{0\, \mu}+
\widehat{\Gamma}^{\gamma W^- W^+}_\mu) \hat{\Delta}_W \Gamma_0\ +\
\Gamma_0 S^{(e)}_0 \Gamma^\gamma_{0\, \mu}\hat{\Delta}_W \Gamma_0\
+\ \Gamma_0 \hat{\Delta}_W \Gamma^\gamma_{0\, \mu} S^{(\mu )}_0 \Gamma_0\, .\
\end{equation}
The PT procedure renders all hatted quantities in the above
expression independent of the gauge-fixing parameter $\xi$;
$\hat{\Delta}_W$ is given in Eq.~(\ref{Dres}).
The final ingredient which enforces the
full $R_\xi$-invariance of the resonant amplitude $\widehat{T}_{1\mu}$,
and allows it to be cast in the form of Eq~(\ref{Twres}),
is a number of Ward identities, satisfied by the PT vertices.
These identities can
be summarized as follows (all momenta flow into the vertex,
{\em i.e.}, $q+p_-+p_+=0$):
\begin{eqnarray}
\frac{1}{e}q^{\mu}\widehat{\Gamma}_{\mu\nu\lambda}^{\gamma W^- W^+}&=&
\widehat{\Pi}^W_{\nu\lambda} (p_-)-\widehat{\Pi}^W_{\nu\lambda} (p_+)\, ,
\label{PTW1}\\
\frac{1}{e}q^{\mu}\widehat{\Gamma}_{\mu\nu}^{\gamma G^- W^+}\ =\
\frac{1}{e}q^{\mu}\widehat{\Gamma}_{\mu\nu}^{\gamma W^- G^+}
&=&
\widehat{\Theta}_\nu (p_-)+\widehat{\Theta}_\nu (p_+)\, ,\label{PTW2}\\
\frac{1}{e}q^{\mu}\widehat{\Gamma}_\mu^{\gamma G^- G^+}&=&
\widehat{\Omega} (p_-)-\widehat{\Omega} (p_+)\, ,\label{PTW3}\\
\frac{1}{e}[p^\nu_-\widehat{\Gamma}_{\mu\nu\lambda}^{\gamma W^- W^+}
-M_W \widehat{\Gamma}_{\mu\lambda}^{\gamma G^- W^+}]
&= &\widehat{\Pi}_{\mu\lambda}^{W}(p_+)-\widehat{\Pi}_{\mu\lambda}^{\gamma}(q)
- \frac{c_w}{s_w}\widehat{\Pi}_{\mu\lambda}^{\gamma Z}(q)\, , \label{PTW4}\\
\frac{1}{e}[p^\lambda_+\widehat{\Gamma}_{\mu\nu\lambda}^{\gamma W^- W^+}
+ M_W \widehat{\Gamma}_{\mu\nu}^{\gamma W^- G^+}]
&= &-\widehat{\Pi}_{\mu\nu}^{W}(p_-)+\widehat{\Pi}_{\mu\nu}^{\gamma}(q)
+\frac{c_w}{s_w}\widehat{\Pi}_{\mu\nu}^{\gamma Z}(q)\, , \label{PTW5}\\
\frac{1}{e}[p^\nu_-\widehat{\Gamma}_{\mu\nu}^{\gamma W^- G^+}
-M_W\widehat{\Gamma}_\mu^{\gamma G^- G^+}] &=&
-\widehat{\Theta}_{\mu}(p_+)\, ,\label{PTW6}\\
\frac{1}{e}[p^\lambda_+\widehat{\Gamma}_{\mu\lambda}^{\gamma G^- W^+}
+ M_W\widehat{\Gamma}_\mu^{\gamma G^- G^+}] &=&
-\widehat{\Theta}_{\mu}(p_-)\, ,\label{PTW7}\\
\frac{1}{e}[p^\nu_-p^\lambda_+
\widehat{\Gamma}_{\mu\nu\lambda}^{\gamma W^- W^+}
+ M_W^2\widehat{\Gamma}_{\mu}^{\gamma G^- G^+} ]
&=& M_W\widehat{\Theta}_\mu (p_+) - M_W\widehat{\Theta}_\mu (p_-)\nonumber\\
&& -p^\lambda_+[\widehat{\Pi}_{\mu\lambda}^{\gamma}(q)
+\frac{c_w}{s_w}\widehat{\Pi}_{\mu\lambda}^{\gamma Z}(q)]\, .\label{PTW9}
\end{eqnarray}
In the derivation of the above equations, we have used the fact
that
\begin{eqnarray}
q^{\mu}\widehat{\Pi}_{\mu\lambda}^\gamma (q)&=&0\, ,\label{trgamgam}\\
q^{\mu}\widehat{\Pi}_{\mu\lambda}^{\gamma Z}(q)&=&0\, ,\label{trgamZ}
\end{eqnarray}
which implies that $\widehat{\Pi}^\gamma_{\mu\nu} (0) =0$ and
$\widehat{\Pi}^{\gamma Z}_{\mu\nu}(0) = 0$.

The one-loop PT self-energy~\cite{JPself} and the one-loop $\gamma WW$
vertex~\cite{JPKP} are respectively given by:
\begin{eqnarray}
\widehat{\Pi}^W_{\mu\nu} (p) & =& \Pi_{\mu\nu}^{W(\xi=1)}(p)\ -
\ 4g^2_w U^{-1}_{W\mu\nu}(p) [ s^2_w I_{W\gamma}(p) + c^2_w I_{WZ} (p) ],
\label{hatPi}   \\
\widehat{\Gamma}^{\gamma W^- W^+}_{\mu\nu\lambda} (q,p_-,p_+) & =&
\Gamma^{\gamma W^- W^+(\xi=1) }_{\mu\nu\lambda}(q,p_-,p_+)\ -\
g_ws_w\Big[
U^{-1\, \alpha}_{\gamma\, \mu}(q) B_{\alpha\nu\lambda}(q,p_-,p_+)\nonumber\\
&&+ U^{-1\, \alpha}_{W \nu}(p_-) B^+_{\mu\alpha\lambda}(q,p_-,p_+)
+ U^{-1\, \alpha}_{W \lambda}(p_+)
B^-_{\mu\nu\alpha}(q,p_-,p_+)\Big] \nonumber\\
&& -\ 2 g_w^2 \Gamma^{\gamma W^- W^+}_{0\mu\nu\lambda}(q,p_-,p_+)\Big[\,
I_{WW}(q) + s^2_w I_{W\gamma}(p_-) \nonumber + s^2_w
I_{W\gamma}(p_+)\nonumber\\
&& + c^2_w I_{WZ}(p_-) + c^2_w I_{WZ} (p_+)\, \Big]\ +\
g_ws_w\Big[ g_{\mu\nu}p_{+\lambda}
{\cal M}^-(q,p_-,p_+)\nonumber\\
&& +\ g_{\mu\lambda}p_{-\nu}{\cal M}^+(q,p_-,p_+) \Big] ,
\label{hatWWgamma}
\end{eqnarray}
where $\Pi_{\mu\nu}^{W(\xi=1)}$~\cite{M&S} and
$\Gamma^{\gamma W^- W^+(\xi=1) }_{\mu\nu\lambda}$~\cite{WWgamma}  are
the conventional one-loop $WW$ self-energy
and $\gamma WW$ coupling, respectively, evaluated in the Feynman gauge,
and the functions $I_{ij}$, $B_{\mu\nu\lambda}$, $B^\pm_{\mu\nu\lambda}$,
and ${\cal M}^\pm$  are defined in Appendix~B.

If we now contract
$\widehat{T}_{1\mu}$ of Eq.~(\ref{Twres})
with $q^{\mu}$, it is elementary to verify, that by virtue
of the Ward identity of Eq.~(\ref{PTW1}),
$q^{\mu}\widehat{T}_{1\mu}=0$. So we conclude that the resonant
amplitude obtained by the PT satisfies both $R_{\xi}$
and $U(1)_{em}$ invariance.

Note finally,
that all PT Green's functions defined thus far satisfy
QED-like
Ward identities (for example, Eqs.~(\ref{PTW1})--(\ref{PTW9})).
This feature not only enforces the $R_{\xi}$ and $U(1)_{em}$
invariance, but it
constitutes a sufficient condition
that our approach admits multiplicative renormalization~\cite{AHKKM}.

Another process that is of particular interest in testing the
electroweak theory at TEVATRON is $QQ'\to e^-\bar{\nu}_e \mu^-\mu^+$;
there, in addition to the $\gamma WW$, the $ZWW$ coupling appears also.
The phenomenological relevance of the $ZWW$ coupling
becomes important as soon as the invariant-mass cut $m(\mu^-\mu^+)
\simeq M_Z$ is imposed. In a similar way, one
can analytically derive the $\widehat{T}_1$ amplitude for this process,
which is more involved due to the presence of $Z\gamma$-mixing
effects~\cite{BC}. As an example, we consider
the g.i.\ amplitude $\widehat{T}^Z_1$, which,
as can be seen from Fig.~9,
does not contain tree-level photonic contributions.
$\widehat{T}^Z_1$ can be cast into the form
\begin{eqnarray}
\widehat{T}^Z_1 &=& \Gamma_0 \hat{\Delta}_W(p_-) (\Gamma^{ZW^- W^+}_0
+\widehat{\Gamma}^{Z W^- W^+})\hat{\Delta}_Z(q)\Gamma^Z_0
\hat{\Delta}_W(p_+) \Gamma_0\nonumber\\
&&+\
\Gamma_0 S^{{}^{(Q)}}_0 \Gamma^Z_0\hat{\Delta}_Z(q)\Gamma^Z_0
\hat{\Delta}_W(p_+) \Gamma_0\
+\ \Gamma^Z_0 S^{{}^{(Q')}}_0 \Gamma_0 \hat{\Delta}_Z(q)\Gamma^Z_0
\hat{\Delta}_W(p_+) \Gamma_0\nonumber\\
&&+\ \Gamma_0 \hat{\Delta}_W(p_-) \Gamma^Z_0
\hat{\Delta}_Z(q)\Gamma^Z_0 S^{(e)}_0 \Gamma_0 \
+\ \Gamma_0 \hat{\Delta}_W(p_-) \Gamma^Z_0 \hat{\Delta}_Z(q)\Gamma_0
S^{(\nu_e)}_0\Gamma^Z_0\nonumber\\
&&+\ \Gamma_0\hat{\Delta}_W(p_-) \Gamma_0 S^{(\nu_\mu )}_0 \Gamma_0
\hat{\Delta}_W(p_+)\Gamma_0,
\label{hatTZ1}
\end{eqnarray}
where $\Gamma^Z_0$ stands for the $Z$ coupling to fermions at the tree
level. The PT Ward identities, which are necessary for maintaining gauge
invariance, are listed in Appendix~C. It should be noted
that the inclusion of the $Z\gamma$ mixing in Eq.~(\ref{hatTZ1})
proceeds in a straightforward way,
since in the PT framework
these additional contributions form
a distinct g.i.\ subset of graphs.
 Indeed, both
$\widehat{\Pi}^\gamma_{\mu\nu} (q)$
and $\widehat{\Pi}^{\gamma Z}_{\mu\nu} (q)$ are by construction
{\it independent}
of the gauge-fixing parameter, and the final gauge cancellations
proceed by virtue of the transversality properties
of $\widehat{\Pi}^\gamma_{\mu\nu} (q)$ and
$\widehat{\Pi}^{\gamma Z}_{\mu\nu} (q)$, as explicitly
stated in Eqs.~(\ref{trgamgam}) and~(\ref{trgamZ}).
By analogy, the Higgs-mixing terms,
which become significant for external
heavy fermions, also form a g.i.\ subset;
possible additional
refinements necessary for their proper inclusion in $\hat{T}_1$
will be studied elsewhere.

\setcounter{equation}{0}
\section{Conclusions}
\indent

We have presented a new g.i.\ approach to resonant transition
amplitudes with external nonconserved currents, based
on the PT method. We have explicitly demonstrated how our analytic
approach bypasses the theoretical difficulties
existing in the present literature,
by considering the resonant processes
$e^-\bar{\nu}_e\to \mu^-\bar{\nu}_\mu$ and $\gamma e^-\to
\mu^-\bar{\nu}_\mu \nu_e $ in the SM, with massive
external charged leptons.
In particular, it has been found that our approach
defines a consistent g.i.\ perturbative expansion of the $S$ matrix,
where singular propagators are regularized by resumming
PT self-energies. Through an explicit proof, particular emphasis
has been put on the fact that the PT resummed propagator does not
shift the complex pole position of the resonant amplitude.
Furthermore, it has been demonstrated that the so-derived propagator
does not give rise to fixed unphysical Landau poles.
The main points of our approach can be summarized as follows:

\begin{itemize}
\item[(i)] The analytic expressions derived with our approach
are, by construction, {\em independent} of the
gauge-fixing parameter, in {\em every} gauge-fixing scheme
($R_{\xi}$ gauges, axial gauges, background field method, {\em etc.}).
In addition, by virtue of the tree-level Ward identities
satisfied by the PT Green's functions,
the $U(1)_{em}$ invariance can be enforced, without
introducing  residual gauge-dependent terms of higher orders.

\item[(ii)] As can be noticed from Section~9 and Appendix~C,
the two- and three-point PT functions satisfy abelian-type
Ward identities. This is a sufficient condition in order that
multiplicative renormalization is admissible within our approach.

\item[(iii)] We treat, {\em on equal
footing}, bosonic and fermionic contributions to the resummed
propagator of the $W$-, $Z$-boson, $t$ quark or other unstable particle.
This feature is highly desirable when confronting the
predictions of
extensions of the SM with data from high energy colliders, such as
the planned Large Hadron Collider at CERN (LHC). Most noticeably,
extra gauge bosons, such as the
$Z'$, $W'$, $Z_R$
predicted in $SO(10)$ or $E_6$ unified models~\cite{GUT},
can have widths predominantly due to bosonic channels; the
same would be true for the standard Higgs boson ($H$)
within the minimal SM, if it turned out to be
heavy.
In such cases it
becomes particularly apparent that prescriptions
based on resumming {\em only}
g.i.\ subsets of fermionic contributions
are bound to be inadequate.

\item[(iv)] The main drawback of using an expansion of the resonant
matrix element in terms of a {\em constant} complex pole is that
this approach introduces space-like threshold terms to all
orders, whereas non-resonant corrections can remove such terms only up
to a given order.
These space-like terms manifest themselves when
the c.m.s.\ energy of the process does
not coincide with the position of the resonant pole. As we showed
in Section 7,
these terms explicitly violate the unitarity of the amplitude.
On the contrary,
our approach avoids this kind of problems by
yielding an energy-dependent complex-pole
regulator.
For instance, for channels below their production threshold, such
residual unitarity-violating terms coming from unphysical absorptive
parts have already been killed by the corresponding kinematic
$\theta$ functions.

\item[(v)] Finally, our approach provides a good high-energy unitarity
behaviour to our amplitude, as the c.m.s.\
energy $s\to \infty$. In fact, far away from the resonance,
the resonant amplitude tends to the usual PT amplitude, showing up
the correct high-energy unitarity limit of the entire tree-level
process.

\end{itemize}

Although more attention has been paid to the unstable $W$ and $Z$
gauge particles, our considerations will also apply to the case
of the heavy top quark discovered recently~\cite{top}.
Our formalism is particularly suited for
a systematic study of the CP properties of the
top quark~\cite{AP} at LHC.
Our method may find important applications in the context of
supersymmetric
theories, especially
when resonant CP effects in the production and
decay of heavy gluinos and scalar quarks are studied~\cite{NP}.
It may also be interesting to consider our g.i.\ approach as an appealing
alternative to the conventional formulation of supergravity theories
in the background field gauges, where, in addition to the regular
Fadeev--Popov ghosts~\cite{FP}, the Nielsen--Kallosh ghosts~\cite{NK}
may appear. Finally, our analysis could be of relevance for the study
of nonperturbative or Coulomb-like phenomena, which may appear
in the production of unstable particles~\cite{GPZ}, and are currently
estimated by using special forms of DS integral equation~\cite{FK,GPZ}.

\vskip1.cm
\noindent
{\bf Acknowledgements.} We wish to thank J.~M.~Cornwall,
K.~Philippides,
A.~Sirlin, R.~Stuart, and D.~Zeppenfeld, for helpful
discussions.
This work was supported in part by the National Science Foundation
Grant No.\ PHY--9313781.

\newpage
\setcounter{section}{0}
\def\theequation{\Alph{section}.\arabic{equation}}
\begin{appendix}
\setcounter{equation}{0}
\section{The structure of the ${\tilde R}^{P}$ terms}
\indent

In order to understand the structure of the ${\tilde R}^{P}$,
we study in detail the three-loop case.
To avoid notational clutter we remove the superscript ``$P$'' from
${\cal V}^{P}$, $V^{P}$, $v^{P}$, ${\cal B}^{P}$, $B^{P}$, and
$b^{P}$.

For $k=3$, Eq.~(\ref{RFormula}) gives
\begin{equation}
{\tilde R}_{3}=
- \left[\Pi_{1}{\cal V}_{2}+ \Pi_{2}{\cal V}_{1}\right], \label{A:R3}
\end{equation}
where we used that $B_{1}={\cal B}_{1}=0$ in the Feynman gauge.

We now proceed to derive Eq.~(\ref{A:R3}).
To that end, we
first express a string with the three $\widehat{\Pi}_1$ self-energies
in terms of conventional strings, and the
necessary pinch contributions.
We have:
\begin{eqnarray}
{\widehat L}_{1}&=&D_{0}
{\widehat{\Pi}}_{1}D_{0}{\widehat{\Pi}}_{1}D_{0}
{\widehat{\Pi}}_{1}D_{0}\nonumber\\
&=& D_{0}
{\left[\Pi_{1}+{\cal V}_{1}D_{0}^{-1}\right]}D_{0}
{\left[\Pi_{1}+{\cal V}_{1}D_{0}^{-1}\right]}D_{0}
{\left[\Pi_{1}+{\cal V}_{1}D_{0}^{-1}\right]}D_{0}
\nonumber\\
&=& D_{0}{\left[\Pi_{1}^{3}D_{0}^{2}+3\Pi_{1}^{2}{\cal V}_{1}D_{0}
+3\Pi_{1}{\cal V}_{1}^{2} + {\cal V}_{1}^{3}D_{0}^{-1}\right]}D_{0}
\nonumber\\
&=& L_{1}+D_{0}{\left[3\Pi_{1}^{2}{\cal V}_{1}D_{0}
+3\Pi_{1}{\cal V}_{1}^{2} + {\cal V}_{1}^{3}D_{0}^{-1}\right]}D_{0}.
\label{A:hatL1}
\end{eqnarray}
In a similar way, we have for the
string containing a ${\widehat{\Pi}}_{1}$ and ${\widehat{\Pi}}_{2}$~:
\begin{eqnarray}
{\widehat L}_{2}&=& 2D_{0}{\widehat{\Pi}}_{1}D_{0}{\widehat{\Pi}}_{2}D_{0}
\nonumber\\
&=& 2D_{0}{\left[\Pi_{1}+{\cal V}_{1}D_{0}^{-1}\right]}D_{0}
{\left[\Pi_{2}+{\cal V}_{2}D_{0}^{-1}
+{\cal B}_{2}D_{0}^{-2}+ {\tilde R}_{2}\right]}D_{0}
\nonumber\\
&=& 2D_{0}\Big[ \Pi_{1}\Pi_{2}D_{0} +
(\Pi_{1}{\cal V}_{2}+\Pi_{2}{\cal V}_{1}-\Pi_{1}{\cal V}_{1}^{2})
-\Pi_{1}^{2}{\cal V}_{1}D_{0}\nonumber\\
&& + (\Pi_{1}{\cal B}_{2}+{\cal V}_{1}{\cal V}_{2})D_{0}^{-1}
+ {\cal V}_{1}{\cal B}_{2}D_{0}^{-2}\Big] D_{0}\nonumber\\
&=& L_{2}+2D_{0}\Big[
(\Pi_{1}{\cal V}_{2}+\Pi_{2}{\cal V}_{1}-\Pi_{1}{\cal V}_{1}^{2})
-\Pi_{1}^{2}{\cal V}_{1}D_{0}
+ (\Pi_{1}{\cal B}_{2}+{\cal V}_{1}{\cal V}_{2})D_{0}^{-1}\nonumber\\
&& + {\cal V}_{1}{\cal B}_{2}D_{0}^{-2}\Big]D_{0}\, ,\label{A:hatL2}
\end{eqnarray}
where we used that ${\tilde R}_{2}=-\Pi_{1}{\cal V}_{1}$.
 From the graphs depicted in Fig.~10, we receive the
propagator-like pinch contributions $L_{3}$, $L_{4}$,
$L_{5}$, $L_{6}$, and $L_{7}$
respectively, given by
\begin{eqnarray}
L_{3} &=& D_{0}\Pi_{1}D_{0}V_{2} =
D_{0}\left[\Pi_{1}V_{2}\right]D_{0},  \label{A:L3}\\
L_{4} &=& D_{0}\Pi_{2}D_{0}V_{1}=
D_{0}\left[\Pi_{2}V_{1}\right]D_{0}\, ,\label{A:L4}\\
L_{5} &=& D_{0}\Pi_{1}D_{0}\Pi_{1}D_{0}V_{1}=
D_{0}\left[\Pi_{1}^{2}V_{1}D_{0}\right]D_{0}\, ,\label{A:L5}\\
L_{6} &=& \frac{1}{4}V_{1}D_{0}\Pi_{1}D_{0}V_{1}=
D_{0}\left[\frac{1}{4}\Pi_{1}V_{1}^{2}\right]D_{0}\, ,\label{A:L6}\\
L_{7} &=& \frac{1}{2}V_{1}D_{0}V_{2}=
D_{0}\left[\frac{1}{2}V_{1}V_{2}D_{0}^{-1}\right]D_{0}\, .\label{A:L7}
\end{eqnarray}

We now add by parts the hatted and unhatted quantities
from Eq.~(\ref{A:R3}) -- (\ref{A:L7}); their difference represents
the contributions one has to
add (and subsequently subtract, as described in Section~2)
in order to convert ``unhatted'' strings
into ``hatted'' strings.
Using the fact that ${\cal V}_{1}=V_{1}$,
$v_{2}=-\frac{3}{4} V_{1}^{2}$,
and
\begin{equation}
{\cal V}_{2}\ =\ V_{2}+v_{2}\ =\ V_{2}-\frac{3}{4} V_{1}^{2}
\end{equation}
we finally have:
\begin{equation}
\sum_{i=1}^{7}L_{i}\ =\ \sum_{j=1}^{2}{\widehat L}_{j}\ +\ D_{0}R_{3}D_{0}
\end{equation}
with
\begin{equation}
R_{3}\ =\ -\left[2\Pi_{1}{\cal B}_{2}+\frac{5}{8}{\cal V}_{1}^{3}+
\frac{3}{2}{\cal V}_{1}{\cal V}_{2}\right]D_{0}^{-1}
- 2{\cal V}_{1}{\cal B}_{2}D_{0}^{-2} -
\left[\Pi_{1}{\cal V}_{2}+ \Pi_{2}{\cal V}_{1}\right]\, .\label{A:R3a}
\end{equation}
{}From Eq.~(\ref{A:R3a}), we obtain
\begin{eqnarray}
v_{3} & = & -\left[2\Pi_{1}{\cal B}_{2}+\frac{5}{8}{\cal V}_{1}^{3}+
\frac{3}{2}{\cal V}_{1}{\cal V}_{2}\right]\, ,\label{A:v3}\\
b_{3} & = & - 2{\cal V}_{1}{\cal B}_{2}\, ,\label{A:b3}
\end{eqnarray}
and
\begin{equation}
{\tilde R}_{3}= - \left[\Pi_{1}{\cal V}_{2}+ \Pi_{2}{\cal V}_{1}\right].
\label{A:R33}
\end{equation}
We notice that all unwanted terms
proportional to $\Pi_{1}^{2}{\cal V}_{1}D_{0}$ have
canceled against each other as they should.
${\tilde R}_{3}$ of Eq.~(\ref{A:R33}) is precisely what
Eq.(\ref{RFormula}) predicts for $k=3$, namely Eq~(A.1).
As we explained in section 3, the $R_{3}$ terms, together
with the $V_{3}$ and $B_{3}$ propagator-like pinch
terms will eventually convert
$\Pi_{3}$ to ${\widehat{\Pi}}_{3}$.

Having gained enough insight on the structure of the
${\tilde R}^{P}$ terms through the study of explicit examples,
we can now generalize our arguments to obtain Eq.(\ref{RFormula}).
For the rest of this Appendix we restore the superscript ``P''

The basic observation is that
the conversion of regular strings of order
$n$ into  ``hatted'' strings
gives rise to ${\tilde R}^{P}_{n}$ terms only when:

\begin{itemize}

\item[(a)] The regular string is of the form
$D_{0}{\Pi}_{k}D_{0}{\Pi}_{\ell}D_{0}$, with $k+\ell=n$, or

\item[(b)] The regular string is of the form
$D_{0}{\Pi}_{k}D_{0}{\Pi}_{\ell}D_{0}{\Pi}_{j}D_{0}$,
with $k+\ell+j=n$.

\end{itemize}

In other words, only
strings with two or three self-energy bubbles give rise to
${\tilde R}^{P}_{n}$ terms. To understand the reason for that,
let us consider a string of order $n$, consisting of
more than three self-energy insertions, {\em i.e.}
\begin{displaymath}
D_0 \Pi_{i_1} D_0\Pi_{i_2}D_0 \Pi_{i_3}D_0\{\cdots \}D_0
\Pi_{i_{k-1}}D_0\Pi_{i_k}D_0\, ,
\end{displaymath}
where $k>3$, and $\sum_{j=1}^{k}(i_{j})=n$.
As discussed in section 3, in order to convert
any of the self-energy bubbles ${\Pi}_{i_{\ell}}$
into ${\widehat{\Pi}}_{i_{\ell}}$
we must supply the appropriate pinch terms of order $i_{\ell}$
(see Eq.~(\ref{GFPT2})), and subsequently subtract them from
other appropriately chosen graphs.
These extra vertex-like pinch terms, of the form
${\cal V}_{i_{\ell}}^{P}D^{-1}_{0}$, cancel one
of the
$D_{0}$ in the string, and give rise to strings of the form
\begin{eqnarray}
&& D_{0}{\Pi}_{i_{1}}D_{0}{\Pi}_{i_{2}}D_{0}\{\cdots \}D_{0}
{\Pi}_{i_{\ell-2}}D_{0}[{\Pi}_{i_{\ell-1}}
{\cal V}_{i_\ell}^{P}]D_{0}\{\cdots \}D_{0}
{\Pi}_{i_{k-1}}D_{0}{\Pi}_{i_{k}}D_{0}\, ,\nonumber\\
&& D_{0}{\Pi}_{i_{1}}D_{0}{\Pi}_{i_{2}}D_{0}\{\cdots \}D_{0}
{\Pi}_{i_{\ell-1}}D_{0}[{\cal V}_{i_\ell}
{\Pi}_{i_{\ell+1}}]D_{0}\{\cdots \}D_{0}
{\Pi}_{i_{k-1}}D_{0}{\Pi}_{i_{k}}D_{0}\, ,\nonumber
\end{eqnarray}
whereas the $D^{-1}_{0}{\cal B}_{\ell}^{P}D^{-1}_{0}$ box-like
terms cancel two of the internal $D_{0}$,
thus leading to a string of the type
\begin{displaymath}
D_{0}{\Pi}_{i_{1}}D_{0}{\Pi}_{i_{2}}D_{0}\{\cdots \}D_{0}
{\Pi}_{i_{\ell-2}}D_{0}[{\Pi}_{i_{\ell-1}}{\cal B}_{i_{\ell}}^{P}
{\Pi}_{i_{\ell+1}}]D_{0}\{\cdots \}D_{0}
{\Pi}_{i_{k-1}}D_{0}{\Pi}_{i_{k}}D_{0}\, .
\end{displaymath}
The terms inside square brackets
in the above expressions contribute
to the quantities ${\tilde R}^{P}_{(i_{\ell-1}+i_\ell)}$,
${\tilde R}^{P}_{(i_\ell+i_{\ell+1})}$, and
${\tilde R}^{P}_{(i_{\ell-1}+i_\ell+i_{\ell+1})}$, respectively.
They will correspondingly be added to the strings
\begin{eqnarray}
&&D_{0}{\Pi}_{i_{1}}D_{0}{\Pi}_{i_{2}}D_{0}\{\cdots \}D_{0}
{\Pi}_{i_{\ell-2}}D_{0}[\Pi_{(i_{\ell-1}+i_\ell)}]D_{0}\{\cdots \}D_{0}
{\Pi}_{i_{k-1}}D_{0}{\Pi}_{i_{k}}D_{0}\, ,\nonumber\\
&&D_{0}{\Pi}_{i_{1}}D_{0}{\Pi}_{i_{2}}D_{0}\{\cdots \}D_{0}
{\Pi}_{i_{\ell-1}}D_{0}[\Pi_{(i_\ell+i_{\ell+1})}]D_{0}\{\cdots \}D_{0}
{\Pi}_{i_{k-1}}D_{0}{\Pi}_{i_{k}}D_{0}\, ,\nonumber\\
&&D_{0}{\Pi}_{i_{1}}D_{0}{\Pi}_{i_{2}}D_{0}\{\cdots \}D_{0}
{\Pi}_{i_{\ell-2}}D_{0}[\Pi_{(i_{\ell-1}+i_\ell+i_{\ell+1})}]
D_{0}\{\cdots \}D_{0}{\Pi}_{i_{k-1}}D_{0}{\Pi}_{i_{k}}D_{0}\, ,\nonumber
\end{eqnarray}
in order to eventually convert
$\Pi_{(i_{\ell-1}+i_\ell)}$,
$\Pi_{(i_\ell+i_{\ell+1})}$, and $\Pi_{(i_{\ell-1}+i_\ell+i_{\ell+1})}$
into
${\widehat {\Pi}}_{(i_{\ell-1}+i_\ell)}$,
${\widehat {\Pi}}_{(i_\ell+i_{\ell+1})}$, and
${\widehat {\Pi}}_{(i_{\ell-1}+i_\ell+i_{\ell+1})}$, respectively.
For example,
the vertex-like piece ${\cal V}_{i_{2}}D_{0}^{-1}$
will give rise to a string of the form
\begin{displaymath}
D_{0}{\Pi}_{i_{1}}D_{0}[{\cal V}_{i_{2}}{\Pi}_{i_{3}}]D_{0}\{\cdots \}D_{0}
{\Pi}_{i_{k-1}}D_{0}{\Pi}_{i_{k}}D_{0}\, ,
\end{displaymath}
which will be
added to the string
$D_{0}{\Pi}_{i_{1}}D_{0}{\Pi}_{(i_{2}+i_{3})}D_{0}\{\cdots \}D_{0}
{\Pi}_{i_{k-1}}D_{0}{\Pi}_{i_{k}}D_{0}$,
as part of the ${\tilde R}^{P}_{(i_{2}+i_{3})}$ term,
whereas the box-like piece ${\cal B}_{i_{2}}D_{0}^{-2}$
will produce a string
\begin{displaymath}
D_{0}[{\Pi}_{i_{1}}{\cal B}_{i_{2}}{\Pi}_{i_{3}}]D_{0}\{\cdots \}D_{0}
{\Pi}_{i_{k-1}}D_{0}{\Pi}_{i_{k}}D_{0}\, ,
\end{displaymath}
which will be added to the string
$D_{0}{\Pi}_{(i_{1}+i_{2}+i_{3})}D_{0}\{\cdots \}D_{0}
{\Pi}_{i_{k-1}}D_{0}{\Pi}_{i_{k}}D_{0}$, as part of the
${\tilde R}^{P}_{(i_{1}+i_{2}+i_{3})}$ terms.

We see therefore that the terms that one needs to add
to a string of order $n$, which contains more than three self-energy
bubbles,
will be absorbed by other strings of the same order, containing
a smaller number of bubbles. Therefore,
the only time that one will obtain terms which must be added to
the string containing the single
self-energy $\Pi_{(i_{1}+i_{2}+ \cdots+i_{k-1}+i_{k})}
=\Pi_{n}$, {\em e.g.}\
they are part of ${\tilde R}^{P}_{n}$,
is if the string has a maximum number
of three self-energies [(a) or (b) above].
A string of type (a) has the form
$L^{(a)}_{(k,n-k)}= D_{0}{\Pi}_{k}D_{0}{\Pi}_{n-k}D_{0}$
and produces a ${\tilde R}_{(k,n-k)}^{P}$ term, given by
${\tilde R}_{(k,n-k)}^{P}= -\frac{1}{2}
[{\Pi}_{k}{\cal V}_{n-k}^{P}+{\cal V}_{k}^{P}{\Pi}_{n-k}]$.
Of course, for every $L^{(a)}_{(k,n-k)}$ there is a
$L^{(a)}_{(n-k,k)}$, giving rise to
${\tilde R}_{(k,n-k)}={\tilde R}_{(n-k,n)}$. So, the total contribution
of strings of type (a) to ${\tilde R}^{P}_{n}$ is
\begin{equation}
{\tilde R}^{P}_{n,(a)}= -\sum_{k=1}^{n}{\tilde R}_{(k,n-k)}^{P}=
-\sum_{k=1}^{n}{\Pi}_{k}{\cal V}_{n-k}^{P}~.
\end{equation}
We now turn to the strings of type (b); their general structure is
$L^{(b)}_{(\ell,n-j,j-\ell)}=D_{0}{\Pi}_{\ell}D_{0}{\Pi}_{n-j}D_{0}
{\Pi}_{j-\ell}D_{0}$, and the contribution to ${\tilde R}^{P}_{n}$
comes from the box-like pinch contribution
${\cal B}_{n-j}$ to the self-energy $\Pi_{n-j}$,
in the middle of the string.
So, the contribution ${\tilde R}_{(\ell,n-j,j-\ell)}^{P}$
from $L^{(b)}_{(\ell,n-j,j-\ell)}$ is given by
${\tilde R}_{(\ell,n-j,j-\ell)}^{P}=-
{\Pi}_{\ell}{\Pi}_{j-\ell}{\cal B}_{n-j}^{P}$,
and the total contribution from strings of type (b) is
\begin{equation}
{\tilde R}^{P}_{n,(b)} =-\sum_{j=1}^{n}\sum_{\ell=1}^{j}
{\tilde R}_{(\ell,n-j,j-\ell)}^{P}= -\sum_{j=1}^{n}\sum_{\ell=1}^{j}
{\Pi}_{\ell}{\Pi}_{j-\ell}{\cal B}_{n-j}^{P}~.
\end{equation}
Clearly,
${\tilde R}^{P}_{n}={\tilde R}^{P}_{n,(a)}+{\tilde R}^{P}_{n,(b)}$,
which is Eq~(\ref{RFormula})~~(for $k=n$).
\setcounter{equation}{0}
\section{One-loop functions}
\indent

Using the sum convention of the momenta  $q+p_1+p_2=0$,
we first define the following useful integrals:
\begin{eqnarray}
I_{ij}(q) & =& \mu^{4-n} \int\frac{d^nk}{i(2\pi)^n}\,
\frac{1}{(k^2-M^2_i)[(k+q)^2-M^2_j]}, \label{B:Iij}\\
J_{ijk}(q,p_1,p_2) & =& \int\frac{d^nk}{i(2\pi)^n}\,
\frac{1}{[(k+p_1)^2-M^2_i][(k-p_2)^2-M^2_j](k^2-M^2_k)}, \label{B:Jijk}\\
J^\mu_{ijk}(q,p_1,p_2) &=& \int\frac{d^nk}{i(2\pi)^n}\,
\frac{k^\mu}{[(k+p_1)^2-M^2_i][(k-p_2)^2-M^2_j](k^2-M^2_k)}, \nonumber\\
&=& p_1^\mu J^-_{ijk}(q,p_1,p_2)\ + \ p_2^\mu J^+_{ijk}(q,p_1,p_2),
\label{B:Jmuijk}
\end{eqnarray}
where the loop integrals are analytically continued in dimensions
$n=4-2\epsilon$.
Armed with the one-loop functions given in
Eqs.~(\ref{B:Iij})--(\ref{B:Jmuijk}),
we can now present the analytic expressions for the functions
$B$, $B^\pm$, and ${\cal M}^\pm$~\cite{JPKP}. They are given by
\begin{eqnarray}
{\cal M}^-(q,p_-,p_+) & = & g^2_w\Big(\frac{s^2_w}{c^2_w}J_{WW\gamma}\ +\
\frac{c^2_w-s^2_w}{2c^2_w}J_{WWZ}\ +\ \frac{1}{2}J_{WWH}\ +\
\frac{1}{2c^2_w}J_{ZHW}\Big), \\
{\cal M}^+(q,p_-,p_+) & = &  -\ {\cal M}^-(q,p_+,p_-), \\
B_{\mu\nu\lambda}(q,p_-,p_+)   & = &  \sum\limits_{V=\gamma,Z}g^2_V
\Bigg\{ g_{\nu\lambda}\Big[p_{-\mu}(J^-_{WWV}-\frac{3}{2}J_{WWV})
+p_{+\mu}(J^+_{WWV}+\frac{3}{2}J_{WWV})\Big]\nonumber\\
&&-g_{\mu\nu}(3p_{-\lambda}J^-_{WWV}+3p_{+\lambda}J^+_{WWV}
+2q_\lambda J_{WWV} )\nonumber\\
&&- g_{\mu\lambda} (3p_{-\nu} J^-_{WWV} +
3p_{+\nu} J^+_{WWV} - 2 q_\nu J_{WWV} )\Bigg\}\, ,\\
B^-_{\mu\nu\lambda}(q,p_-,p_+) & = & \sum\limits_{V=\gamma,Z}g^2_V
\Bigg\{ g_{\nu\lambda}\Big[ 3p_{-\mu}(J^-_{WWV} +J_{WWV})+
p_{+\mu}(3J^+_{WWV}-2J_{WWV})\Big]\nonumber\\
&&+g_{\mu\lambda}\Big[ p_{-\nu}(3J^-_{WWV}+J_{WWV})+
3p_{+\nu}J^+_{WWV}-2q_\nu J_{WWV}\Big]\nonumber\\
&&-g_{\nu\mu}\Big[ p_{-\lambda}(J^-_{WWV}+2J_{WWV})
+p_{+\lambda}J^+_{WWV}-2q_\lambda J_{WWV}\Big]\, \Bigg\}\, ,\\
B^+_{\mu\nu\lambda}(q,p_-,p_+) & = & -\ B^-_{\mu\lambda\nu}(q,p_+,p_-) ,
\end{eqnarray}
where the coupling constants have been abbreviated by
$g_\gamma=g_w s_w=e$ and $g_Z = g_w c_w$, and the arguments of the
functions $J$, $J_{ijk}$, and $J^\pm_{ijk}$ should be evaluated at
$(q,p_-,p_+)$.

The one-loop functions $I_{ij}$, $J_{ijk}$, and $J^\mu_{ijk}$
defined in Eqs.~(\ref{B:Iij})--(\ref{B:Jmuijk}) are closely
related to the Passarino--Veltman~\cite{PV} integrals. In this way,
if we adopt the Minskowskian  metric $g^{\mu\nu} =\mbox{diag}(1,-1,-1,-1)$
in our conventions, very similar to Ref.~\cite{BAK}, we can
make the following identifications:
\begin{eqnarray}
I_{ij} (q) & = & \frac{1}{16\pi^2} (1+2\epsilon\ln 2\pi\mu)\,
B_0(q^2,M^2_i,M^2_j)\, ,\\
J_{ijk}(q,p_1,p_2) & = & -\, \frac{1}{16\pi^2}\,
C_0(p_1^2,q^2,p_2^2,M^2_k,M^2_i,M^2_j)\, ,\\
J^\mu_{ijk}(q,p_1,p_2) & = & -\, \frac{1}{16\pi^2}\, \Big[
p_1^\mu C_{11}(p_1^2,q^2,p_2^2,M^2_k,M^2_i,M^2_j)\,
+\, q^\mu C_{12}(p_1^2,q^2,p_2^2,M^2_k,M^2_i,M^2_j)\Big]\, .\nonumber\\
&&\label{B:PVJmu}
\end{eqnarray}
 From Eq.~(\ref{B:PVJmu}), it is then easy to derive that
\begin{eqnarray}
J^-_{ijk}(q,p_1,p_2) & = & \hspace{-0.3cm}- \frac{1}{16\pi^2}\Big[
C_{11}(p_1^2,q^2,p_2^2,M^2_k,M^2_i,M^2_j) -
C_{12}(p_1^2,q^2,p_2^2,M^2_k,M^2_i,M^2_j)\Big],\qquad\\
J^+_{ijk}(q,p_1,p_2) & = & \frac{1}{16\pi^2}\,
C_{12}(p_1^2,q^2,p_2^2,M^2_k,M^2_i,M^2_j).
\end{eqnarray}

\setcounter{equation}{0}
\section{Ward identities for the {\em ZWW} vertex}
\indent

Using the PT, one can derive all the relevant Ward identities
related to the $ZWW$ vertex, which warrant an analytic
g.i.\ result. These identities are listed below
\begin{eqnarray}
\frac{1}{g_wc_w}[q^{\mu}\widehat{\Gamma}_{\mu\nu\lambda}^{Z W^- W^+}-
M_Z\widehat{\Gamma}_{\nu\lambda}^{G^0 W^- W^+}] &=&
\widehat{\Pi}^W_{\nu\lambda} (p_-)-\widehat{\Pi}^W_{\nu\lambda} (p_+)\, ,
\label{PTZ1}\\
\frac{1}{g_wc_w}[q^{\mu}\widehat{\Gamma}_{\mu\nu}^{Z W^- G^+} -
M_Z\widehat{\Gamma}_{\nu}^{G^0 W^- G^+}]
&=& \widehat{\Theta}_\nu (p_-)+\widehat{\Theta}_\nu (p_+)\, ,\label{PTZ2}\\
\frac{1}{g_wc_w}[q^{\mu}\widehat{\Gamma}_\mu^{Z G^- G^+}
- M_Z\widehat{\Gamma}^{G^0 G^- G^+} ]
&=& \widehat{\Omega} (p_-)-\widehat{\Omega} (p_+)\, ,\label{PTZ3}\\
\frac{1}{g_wc_w}[p^\nu_-\widehat{\Gamma}_{\mu\nu\lambda}^{Z W^- W^+}
-M_W \widehat{\Gamma}_{\mu\lambda}^{Z G^- W^+}]
&= &\widehat{\Pi}_{\mu\lambda}^W(p_+)-\widehat{\Pi}_{\mu\lambda}^Z(q)
- \frac{s_w}{c_w}\widehat{\Pi}_{\mu\lambda}^{Z \gamma}(q)\, , \label{PTZ4}\\
\frac{1}{g_wc_w}[p^\lambda_+\widehat{\Gamma}_{\mu\nu\lambda}^{Z W^- W^+}
+ M_W \widehat{\Gamma}_{\mu\nu}^{Z W^- G^+}]
&= &-\widehat{\Pi}_{\mu\nu}^W(p_-)+\widehat{\Pi}_{\mu\nu}^Z(q)
+\frac{s_w}{c_w}\widehat{\Pi}_{\mu\nu}^{Z \gamma}(q)\, , \label{PTZ5}\\
\frac{1}{g_wc_w}[p^\nu_-\widehat{\Gamma}_{\mu\nu}^{Z W^- G^+}
-M_W\widehat{\Gamma}_\mu^{Z G^- G^+}] &=& -\widehat{\Theta}_{\mu}(p_+)\, ,
\label{PTZ6}\\
\frac{1}{g_wc_w}[p^\lambda_+\widehat{\Gamma}_{\mu\lambda}^{Z G^- W^+}
+M_W\widehat{\Gamma}_\mu^{Z G^- G^+}] &=& -\widehat{\Theta}_{\mu}(p_-)\, ,
\label{PTZ7}\\
\frac{1}{g_wc_w}[p^\nu_-\widehat{\Gamma}_{\nu\lambda}^{G^0 W^- W^+}
-c_w q^\mu\widehat{\Gamma}_{\mu\lambda}^{Z G^- W^+} ] &=&
c_w \widehat{\Theta}_\lambda(p_-) + c_w\widehat{\Theta}_\lambda(p_+)+
\widehat{\Pi}_\lambda^{Z G^0}(q)\, , \label{PTZ8}\\
\frac{1}{g_wc_w}[p^\lambda_+\widehat{\Gamma}_{\nu\lambda}^{G^0 W^- W^+}
-c_w q^\mu\widehat{\Gamma}_{\mu\nu}^{Z W^- G^+} ] &=&
c_w \widehat{\Theta}_\nu (p_-) + c_w\widehat{\Theta}_\nu (p_+) +
\widehat{\Pi}_\nu^{Z G^0}(q)\, , \label{PTZ9}\\
\frac{1}{g_wc_w}[p^\nu_-p^\lambda_+
\widehat{\Gamma}_{\mu\nu\lambda}^{Z W^- W^+}
+ M^2_W\widehat{\Gamma}_{\mu}^{Z G^- G^+}] &=& M_W\widehat{\Theta}_\mu (p_+)-
M_W\widehat{\Theta}_\mu (p_-)\nonumber\\
&&-\frac{1}{2} (p_+-p_-)^\lambda \Big[ \widehat{\Pi}^Z_{\mu\lambda}(q)+
\frac{s_w}{c_w}\widehat{\Pi}_{\mu\lambda}^{Z\gamma} (q) \Big].
\label{PTZ}
\end{eqnarray}

The PT three-point function for the $ZWW$ coupling is related to
the conventional vertex in the Feynman gauge via the following
expression:
\begin{eqnarray}
\widehat{\Gamma}^{Z W^- W^+}_{\mu\nu\lambda} (q,p_-,p_+) & =&
\Gamma^{Z W^- W^+(\xi=1) }_{\mu\nu\lambda}(q,p_-,p_+)\ -\
g_wc_w\Big[
U^{-1\, \alpha}_{Z \mu}(q) B_{\alpha\nu\lambda}(q,p_-,p_+)\nonumber\\
&&+ U^{-1\, \alpha}_{W\, \nu}(p_-) B^+_{\mu\alpha\lambda}(q,p_-,p_+)
+ U^{-1\, \alpha}_{W \lambda}(p_+)
B^-_{\mu\nu\alpha}(q,p_-,p_+)\Big] \nonumber\\
&& -\ 2 g_w^2 \Gamma^{Z W^- W^+}_{0\mu\nu\lambda}(q,p_-,p_+)\Big[\,
I_{WW}(q) + s^2_w I_{W\gamma}(p_-) \nonumber + s^2_w
I_{W\gamma}(p_+)\nonumber\\
&& + c^2_w I_{WZ}(p_-) + c^2_w I_{WZ} (p_+)\, \Big]\ +\
g_wc_w\Big[ M^2_Wg_{\mu\nu}p_{+\lambda}
{\cal M}^-(q,p_-,p_+)\nonumber\\
&& +\ M^2_Wg_{\mu\lambda}p_{-\nu}{\cal M}^+(q,p_-,p_+)\ +\
M^2_Zq_\mu g_{\nu\lambda}{\cal M}(q,p_-,p_+)\Big] .
\label{hatWWZ}
\end{eqnarray}
In Eq.~(\ref{hatWWZ}), $\Gamma^{Z W^- W^+(\xi=1) }_{\mu\nu\lambda}$
is the conventional one-loop $ZWW$ vertex calculated in the Feynman
gauge. The loop functions $I_{ij}$, $B^\pm$, ${\cal M}^\pm$ are given
in Appendix B, except of ${\cal M}$. The analytic result for the latter
may be obtained by
\begin{equation}
{\cal M}(q,p_-,p_+)\ =\ \frac{1}{2}\, g^2_w \Big[ J_{HZW}(q,p_-,p_+)\ +\
J_{ZHW}(q,p_-,p_+)\Big].
\end{equation}

\end{appendix}
\newpage


\newpage

\centerline{\bf\Large Figure Captions }
\vspace{-0.2cm}
\newcounter{fig}
\begin{list}{\bf\rm Fig. \arabic{fig}: }{\usecounter{fig}
\labelwidth1.6cm \leftmargin2.5cm \labelsep0.4cm \itemsep0ex plus0.2ex }

\item The PT decomposition of the process $e^-\bar{\nu}_e\to
\mu^-\bar{\nu}_\mu$ (the arrow of time shows downwards).

\item The PT method applied to the scattering $q\bar{q}\to q'\bar{q}'$
at the two-loop QCD order.

\item Two-loop PT contributions to the gluon vacuum polarization.

\item Typical two-loop vertex and box graphs giving PT contributions
to the two-loop PT self-energy

\item The propagator-like part $\widehat{T}_1$ of the transition element for
the process
$e^-\bar{\nu}_e\to \mu^-\bar{\nu}_\mu$ at the two-loop electroweak order.

\item The process $l\bar{\nu}_l\to H W^-$ in an arbitrary $R_\xi$ gauge

\item The one-loop absorptive graphs of the reaction
$e^-\bar{\nu}_e\to \mu^-\bar{\nu}_\mu$ involving the on-shell
intermediate bosons $W^-$ and $H$ (the arrow of time shows downwards).
Feynman lines with Goldstone bosons are not displayed.

\item The process $e^-\gamma \to \mu^-\bar{\nu}_\mu \nu_e$. The bubbles
denote PT self-energies and three-point functions. Goldstone boson
lines are not shown.

\item The process $QQ'\to \mu^+\mu^- e^-\bar{\nu}_e$, where $Z\gamma$-mixing
effects and other photonic contributions are not shown. Crossed $Z$-boson
exchange graphs are also implied.

\item Structures of Feynman graphs responsible for the vanishing
of the shift of the pole at the three-loop case ---see, also, Appendix A.

\end{list}

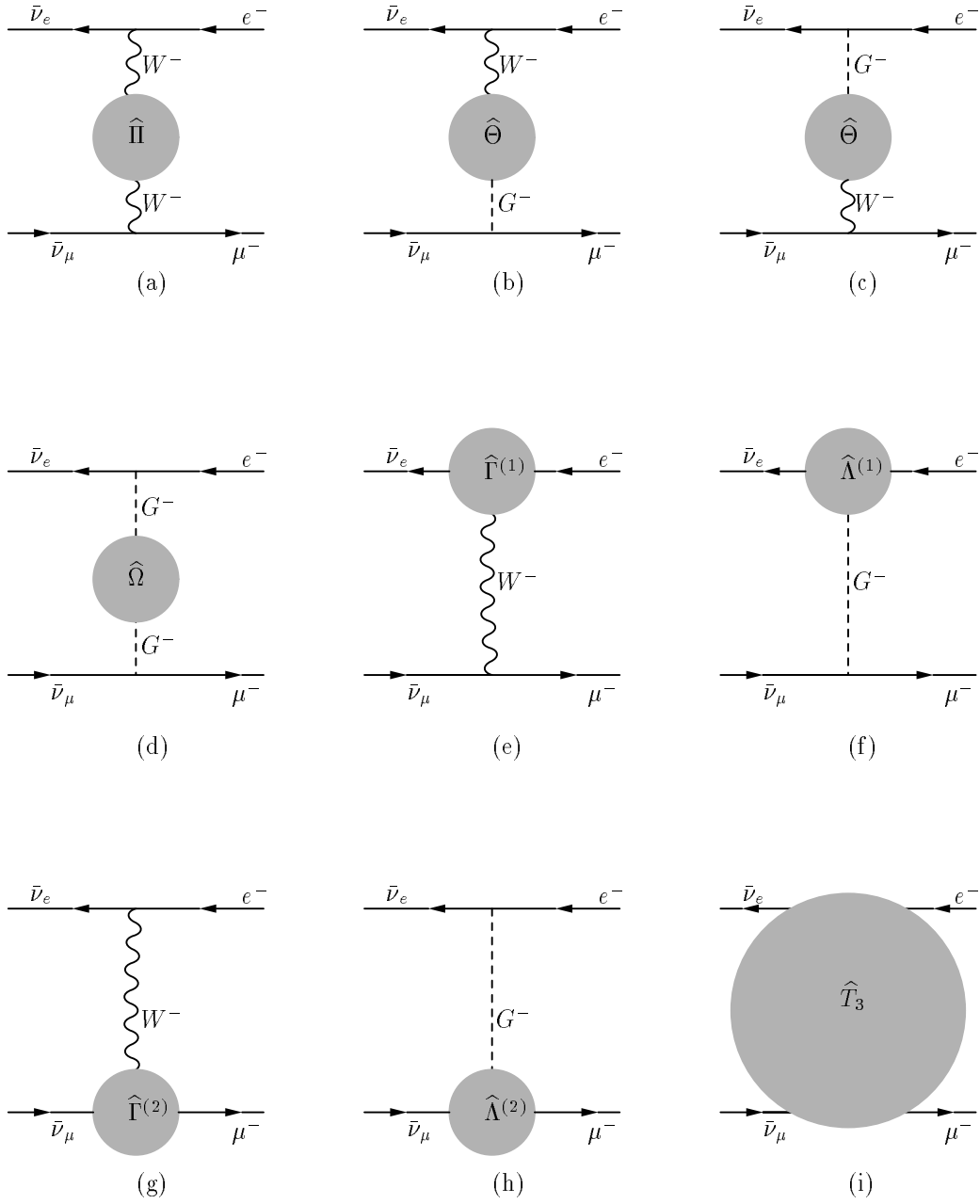
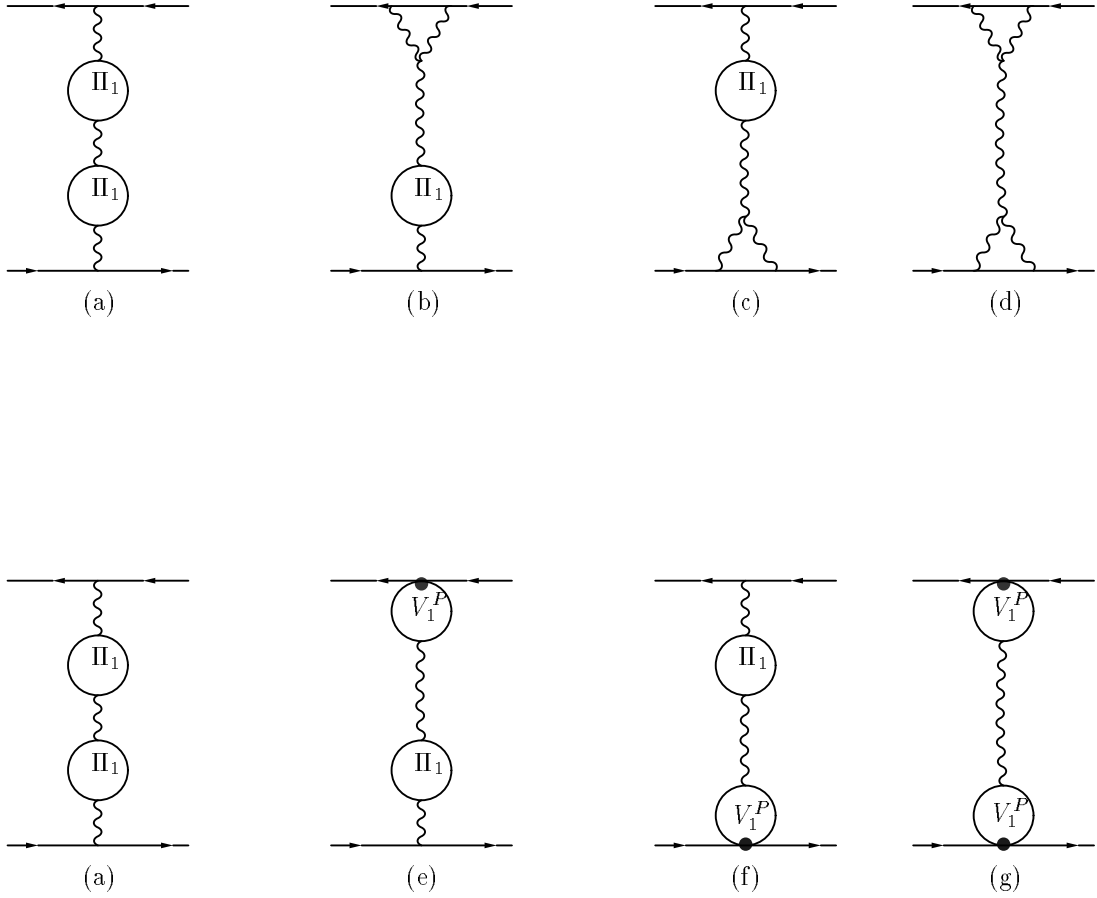
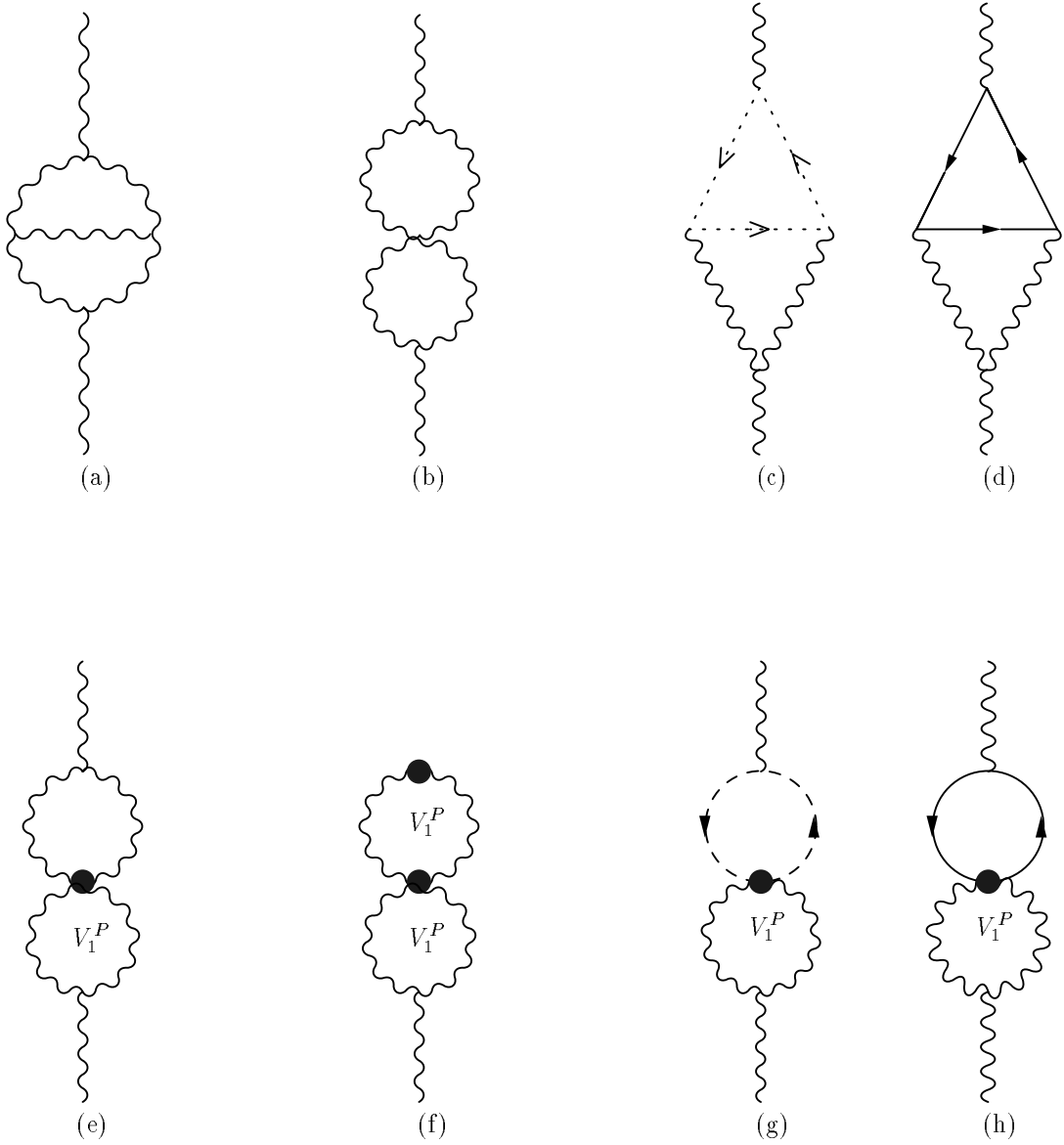
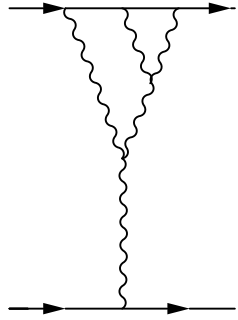
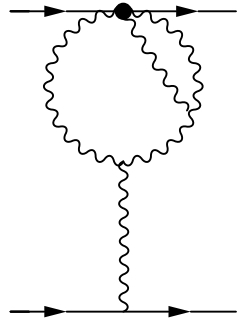
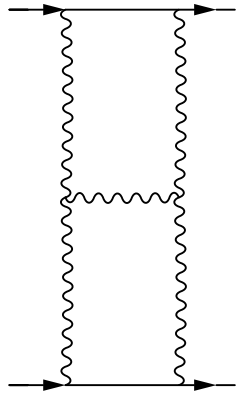
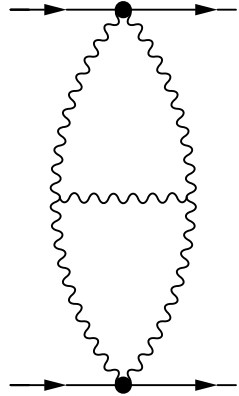
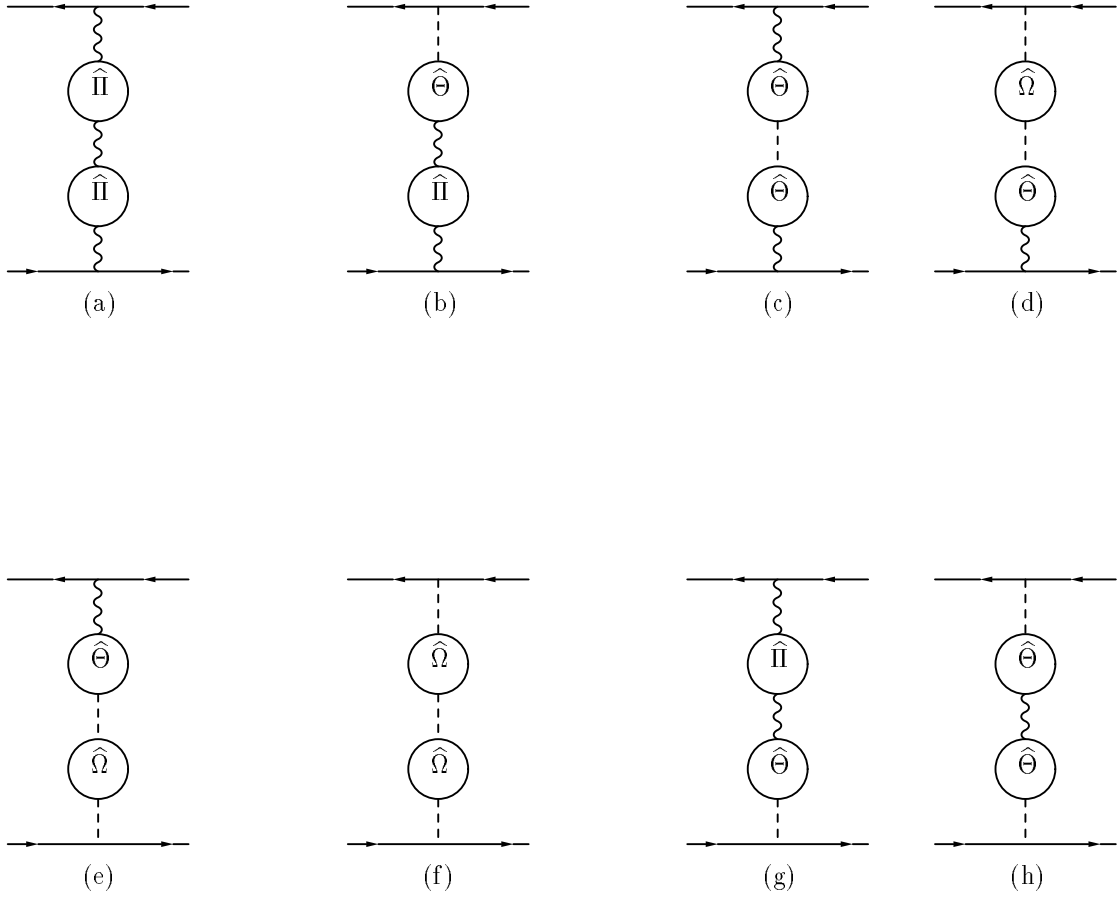
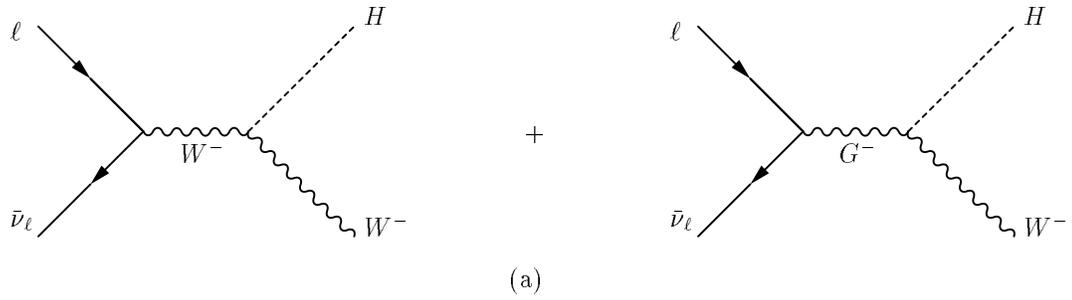
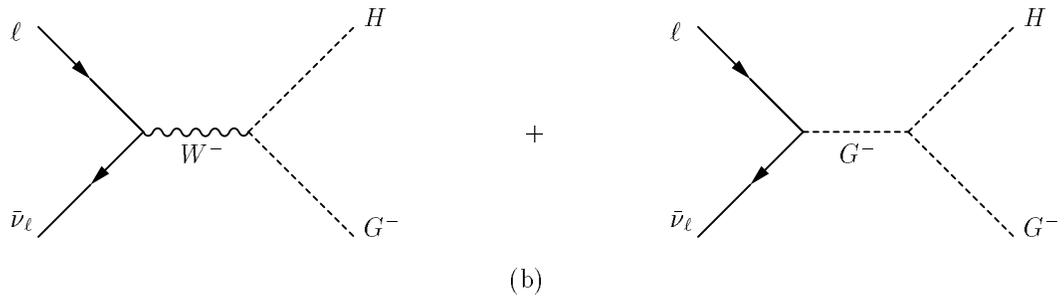
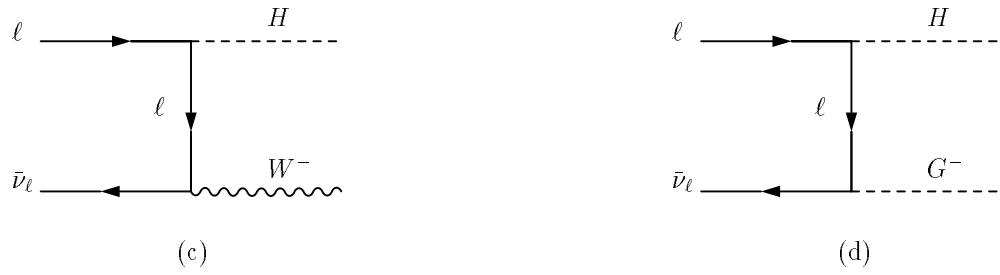
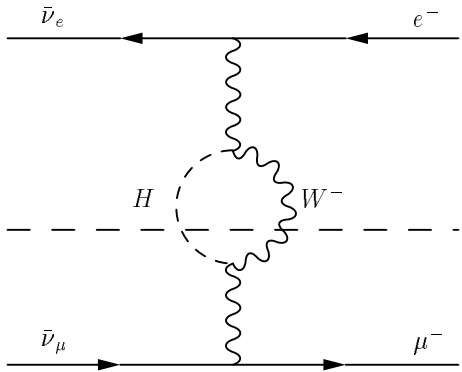
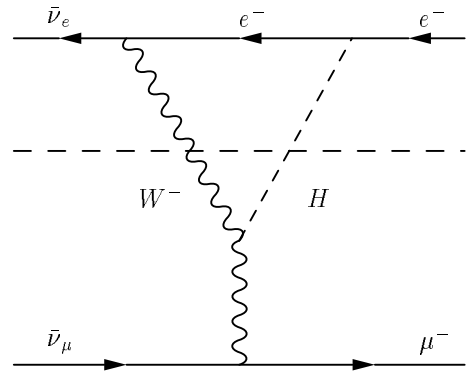
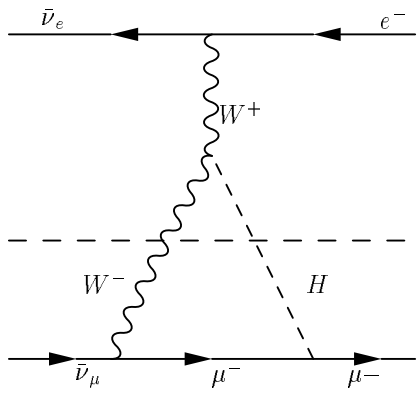
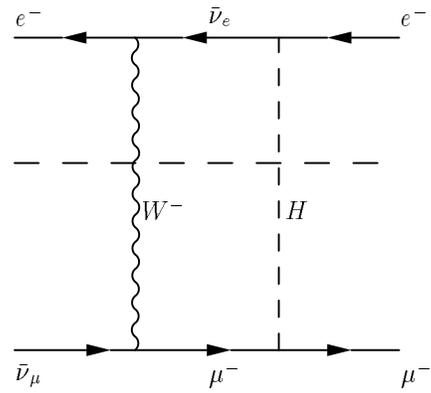
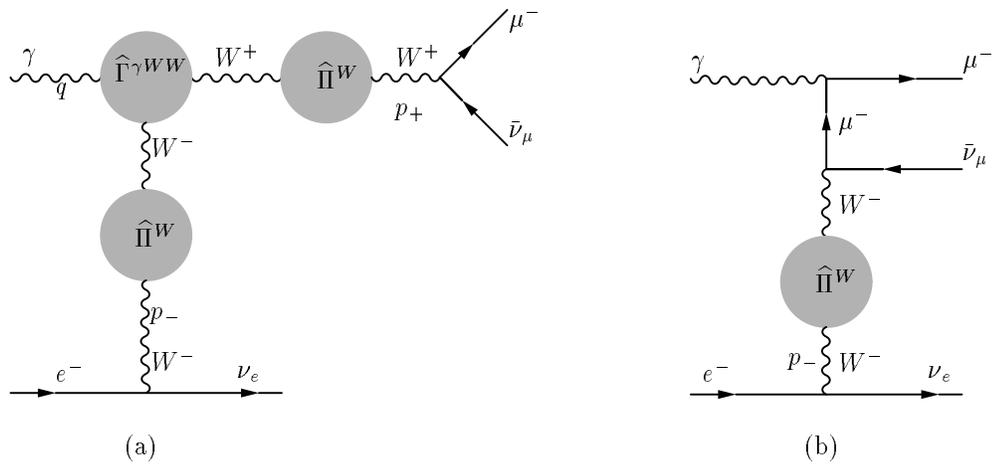
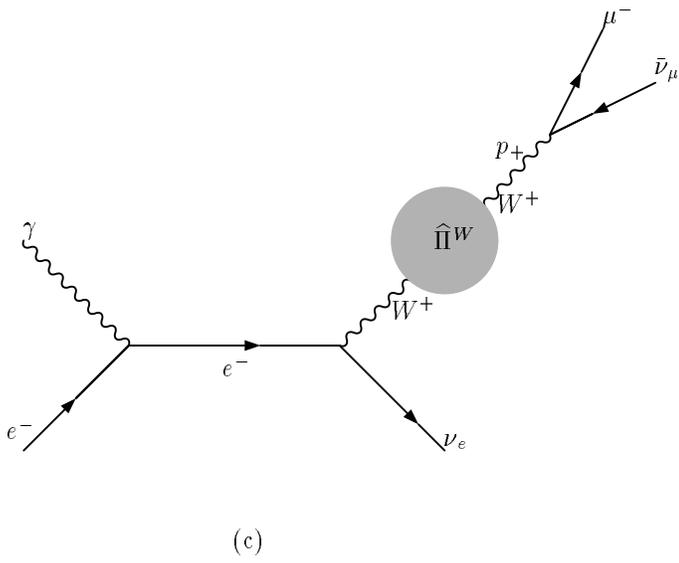
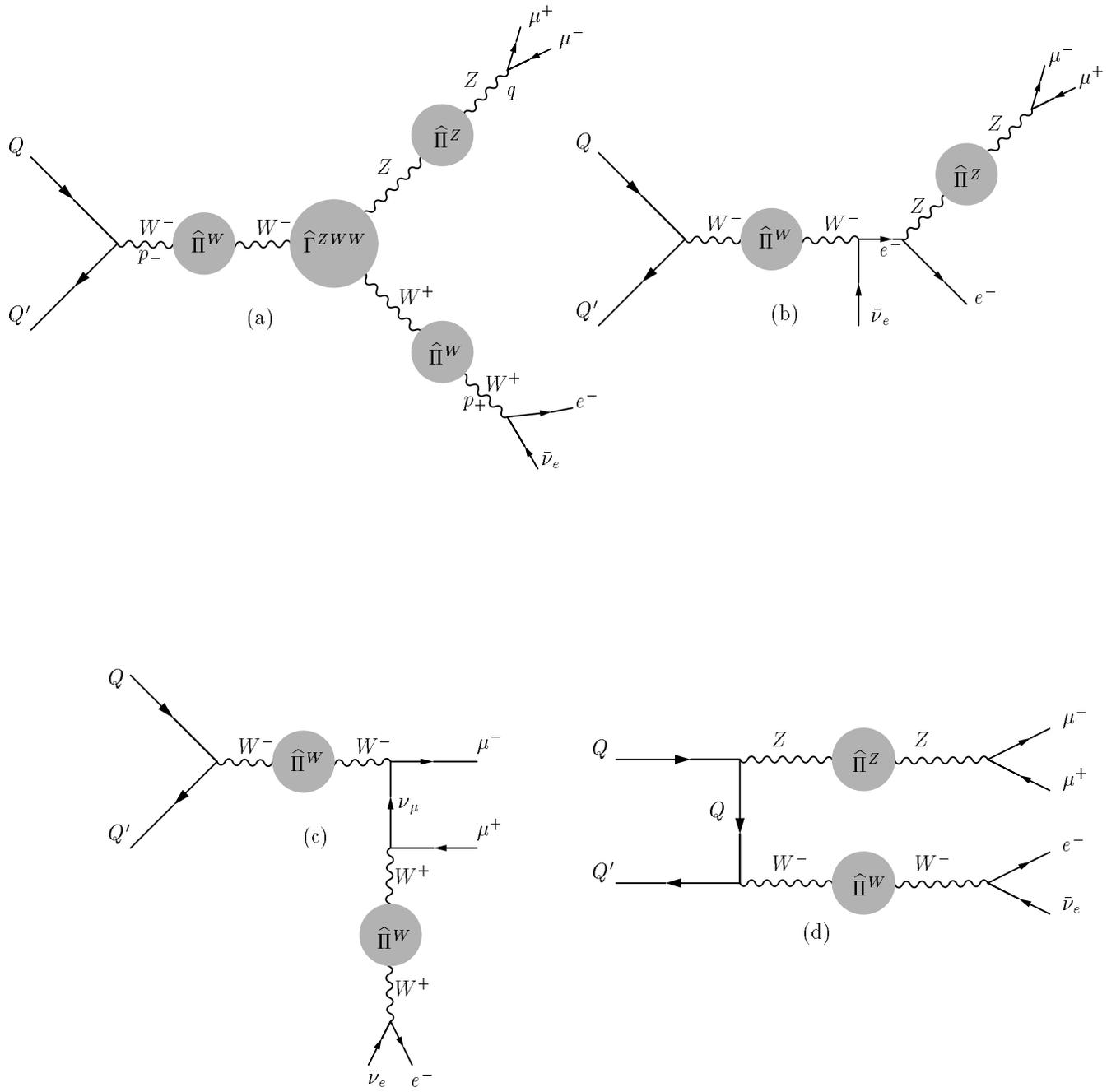
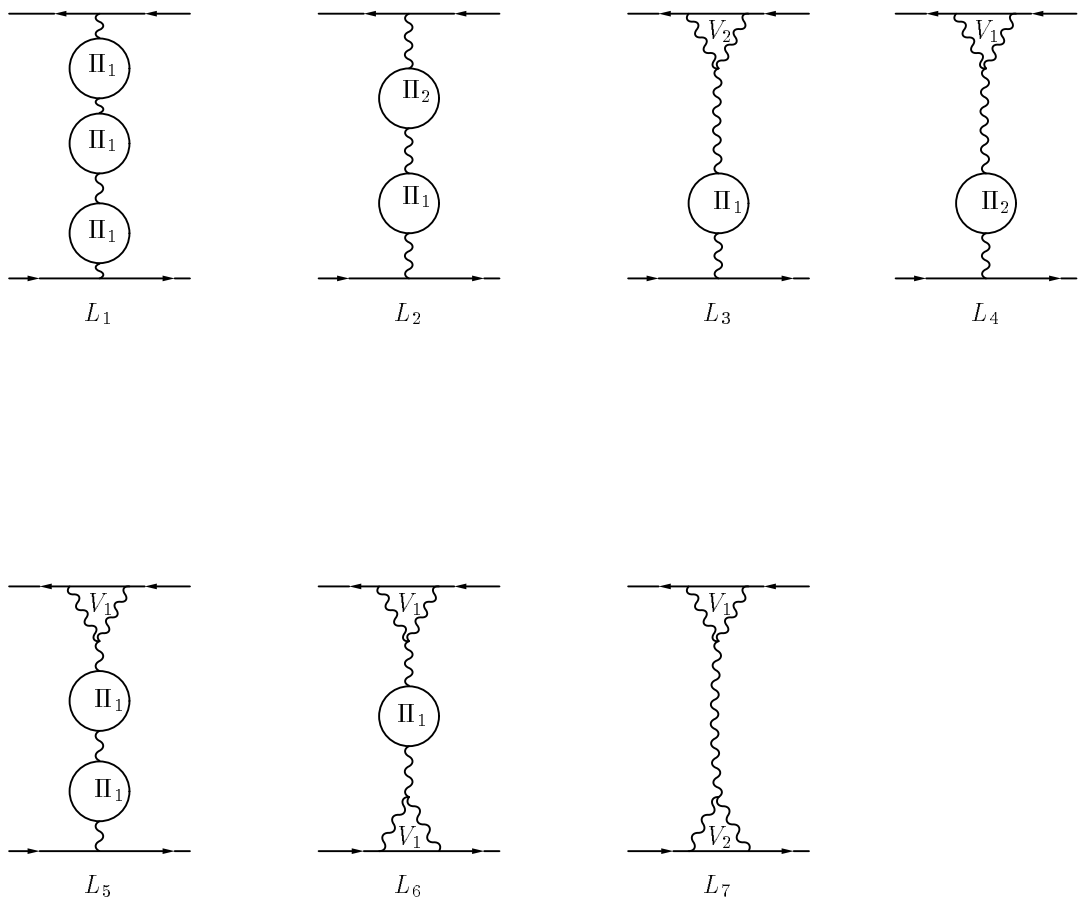
\end{document}